\documentclass[]{aa}
\usepackage{graphics}
\usepackage{graphicx}
\usepackage{txfonts}
\usepackage{natbib}
\bibpunct{(}{)}{;}{a}{}{,}

\def\apj{ApJ}                 
                
\def\apjs{ApJS}

\def\aap{A\&A}

\def\mnras{MNRAS}

\def\zap{ZAp}

\def\nphysa{Nucl.~Phys.~A}

\def\msol{M$_{\odot}$}
\def\lsol{L$_{\odot}$}
\begin{document}      

   \title{Asteroseismic signatures of helium gradients in late F-type stars}

        \titlerunning{Asteroseismic signatures of helium gradients}  
 
   \author{
M. Castro \ 
\and S. Vauclair 
}
 
   \offprints{M. Castro}

   \institute{Laboratoire d'Astrophysique de Toulouse et Tarbes - UMR 5572 - Universit\'e Paul Sabatier Toulouse III - CNRS, 14, av. E. Belin, 31400 Toulouse, France} 

\mail{mcastro@ast.obs-mip.fr}
   
\date{Received \rule{2.0cm}{0.01cm} ; accepted \rule{2.0cm}{0.01cm} }

\authorrunning{M.Castro \& S.Vauclair}

\abstract
 {Element diffusion is expected to occur in all kinds of stars : according to the relative effect of gravitation and radiative acceleration, they can fall or be pushed up in the atmospheres. Helium sinks in all cases, thereby creating a gradient at the bottom of the convective zones. This can have important consequences for the sound velocity, as has been proved in the sun with helioseismology.}
 {We investigate signatures of helium diffusion in late F-type stars by asteroseismology.}
 {Stellar models were computed with different physical inputs (with or without element diffusion) and iterated in order to fit close-by evolutionary tracks for each mass. The theoretical oscillation frequencies were computed and compared for pairs of models along the tracks. Various asteroseismic tests (large separations, small separations, second differences) were used and studied for the comparisons.}
 {The results show that element diffusion leads to changes in the frequencies for masses larger than 1.2 \msol. In particular the helium gradient below the convective zone should be detectable through the second differences.}
 {}


\maketitle
                                                                                                                                               
\section{Introduction} 

Among the different physical processes that occur in stellar interiors, element diffusion plays an important role. Under the oppositing effects of gravitation and thermal settling on the one hand, and radiative acceleration on the other hand, the various elements undergo relative separation. This process, which was described by \citet{michaud70} to account for the abundance anomalies observed in chemically peculiar A stars, is now recognized to occur in all kinds of stars. 

\citet{vauclair04} described in terms of the so-called ``second differences'' an asteroseismic signature of helium diffusion in main-sequence A stars (1.6 \msol \ and 2.0 \msol). \citet{theado05} presented a first discussion of asteroseismic signatures of diffusion in stars between 1.1 and 1.3 \msol. In the present paper we show evidence of differences in the internal structure of stars between 1.1 and 1.45 \msol \ due to element diffusion. We compute evolutionary tracks of stellar models with similar external parameters (luminosities, effective temperatures and chemical compositions), with or without helium diffusion. We then study asteroseismic tests, particularly the signatures of helium gradients below the outer convective zones. 
  
In Section 2, we present the model computations and calibrations. The asteroseismic tests of internal structure between models with and without diffusion are discussed in Section 3. The helium gradients and their consequences for the second differences are studied in Section 4. Section 5 gives the discussion and conclusion.

\section{Calibration and characteristics of the models}

We computed models with the Toulouse-Geneva Evolutionary Code (TGEC) \citep{richard96}, for masses from 1.1 to 1.45 \msol, respectively noted M1.1 to M1.45 (see Table \ref{tab:models}). The physical inputs and parameters are :
\begin{itemize}
\item[-] Equation of state : MHD \citep{dappen92},
\item[-] Opacities : OPAL \citep{iglesias96} completed by the \citet{alex94} low temperature opacities,
\item[-] Nuclear reaction rates : NACRE compilation of nuclear reaction rates \citep{angulo99} with the Bahcall screening routine,
\item[-] Convection treatment : mixing-length theory \citep{bohm58},
\item[-] Diffusion : diffusion coefficients computed as in \citet{paquette86}.
\end{itemize}

For a given stellar mass, we computed two series of models : with and without element diffusion. No mixing process or mass loss is taken into account. The series were calibrated to obtain evolutionary tracks very close in the HR diagram, such that, for the same luminosity, their temperature differences are never larger than 60 K. This calibration is achieved by adjusting two free parameters of the stellar evolution code : the initial helium abundance Y$_0$ and the mixing length parameter $\alpha$. We used for all the homogeneous models the same values of the free parameters $\alpha$ and Y$_0$. For each mass, models including diffusion are then calibrated to have evolutionary tracks very close to the homogeneous ones. The calibration parameters are presented in Table \ref{tab:calib}.

\begin{table*}
\caption{Calibration parameters of the models; $\alpha$: mixing length parameter; Y$_0$: initial helium abundance}
\label{tab:calib}
\begin{center}
\begin{tabular}{c|ccc}
\hline
Models & Mass (\msol) & $\alpha$ & Y$_0$ \\
\hline \hline
M1.1-hom & 1.1 & 1.75 & 0.268 \\
M1.1-dif & 1.1 & 1.84 & 0.267 \\
\hline
M1.2-hom & 1.2 & 1.75 & 0.268 \\
M1.2-dif & 1.2 & 1.88 & 0.268 \\
\hline
M1.3-hom & 1.3 & 1.75 & 0.268 \\
M1.3-dif & 1.3 & 1.99 & 0.268 \\ 
\hline
M1.35-hom & 1.35 & 1.75 & 0.268 \\
M1.35-dif & 1.35 & 1.99 & 0.268 \\
\hline
M1.45-hom & 1.45 & 1.75 & 0.268 \\
M1.45-dif & 1.45 & 1.88 & 0.278 \\
\hline
\end{tabular}
\end{center}
\end{table*}

\begin{figure}
\begin{center}
\includegraphics[angle=0,totalheight=\columnwidth,width=\columnwidth]{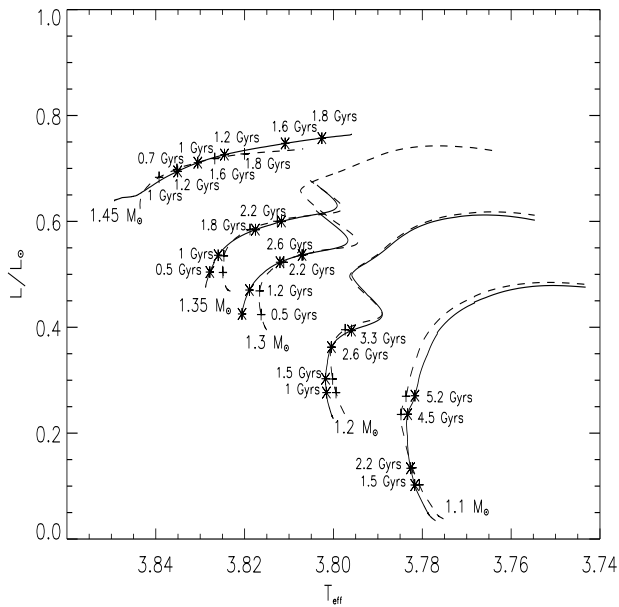}
\end{center}
\caption{Evolutionary tracks of the 1.1 to 1.45 \msol \ models. Dashed lines: homogeneous models, solid lines: models including microscopic diffusion}
\label{HR}
\end{figure}

Figure \ref{HR} displays the computed evolutionary tracks. For each mass the tracks obtained with and without diffusion lie very close to each other. Furthermore, up to 1.35 \msol \ the age evolution is similar in the two cases, so that models with the same age lie close together in the HR diagram. Regarding the uncertainties, we can consider that these models are observationaly identical, i.e. with the same effective temperatures, luminosities and chemical compositions. This is no longer the case for larger masses : along track M1.45, of 1.45 \msol, the ages for models with diffusion are quite different to those without diffusion. Indeed, for masses greater than 1.3 \msol, more than half of the energy production is dominated by the CNO cycle, which is more sensitive to the central temperature than the pp chains. Since models with diffusion have a larger initial helium fraction than models without diffusion, the former are more evolved for the same age. 

Table \ref{tab:models} gives the characteristics of the models with and without diffusion for each mass and different ages. The calibrated models do not have strictly identical values of the surface parameters but they are very close, except for the case M1.45 for which models with similar parameters have different ages. 

\begin{table*}
\caption{Physical characteristics of the models; L/\lsol : luminosity in solar unity; $\vert\Delta$L$\vert$/L : relative difference of luminosity between models with and without diffusion; T$_{eff}$ : effective temperature; $\vert\Delta$T$\vert$/T : relative difference of effective temperature between models with and without diffusion; X$_S$ and Y$_S$: surface hydrogen and helium mass fractions}
\label{tab:models}
\begin{center}
\begin{tabular}[h!]{c|ccccccc} 
\hline 
Models & Ages (Gyr) & L/\lsol & $\vert\Delta$L$\vert$/L & T$_{eff}$ (K) & $\vert\Delta$T$\vert$/T & X$_S$ & Y$_S$ \\
\hline \hline
M1.1-hom & 1.5 & 1.26637 & & 6034.90 & & 0.7130 & 0.2680 \\
 & 2.2 & 1.35947 & & 6056.06 & & 0.7130 & 0.2680 \\
 & 4.5 & 1.71937 & & 6092.56 & & 0.7130 & 0.2680 \\
 & 5.2 & 1.86054 & & 6076.73 & & 0.7130 & 0.2680 \\
\hline
M1.1-dif & 1.5 & 1.26488 & 0.12\% & 6049.23 & 0.24\% & 0.7298 & 0.2517 \\
 & 2.2 & 1.36013 & 0.05\% & 6062.48 & 0.11\% & 0.7379 & 0.2439 \\
 & 4.5 & 1.72119 & 0.11\% & 6071.28 & 0.35\% & 0.7604 & 0.2222 \\
 & 5.2 & 1.86513 & 0.25\% & 6048.67 & 0.46\% & 0.7667 & 0.2162 \\
\hline \hline
M1.2-hom & 1 & 1.89069 & & 6301.30 & & 0.7130 & 0.2680 \\
 & 1.5 & 2.00586 & & 6312.19 & & 0.7130 & 0.2680 \\
 & 2.6 & 2.30632 & & 6316.55 & & 0.7130 & 0.2680 \\
 & 3.3 & 2.48519 & & 6271.48 & & 0.7130 & 0.2680 \\
\hline
M1.2-dif & 1 & 1.89099 & 0.016\% & 6332.72 & 0.50\% & 0.7335 & 0.2482 \\
 & 1.5 & 2.00812 & 0.11\% & 6335.34 & 0.37\% & 0.7438 & 0.2382 \\
 & 2.6 & 2.30420 & 0.09\% & 6315.97 & 0.01\% & 0.7661 & 0.2168 \\
 & 3.3 & 2.47611 & 0.37\% & 6250.72 & 0.33\% & 0.7745 & 0.2087 \\
\hline
 \hline
M1.3-hom & 0.5 & 2.65461 & & 6551.19 & & 0.7130 & 0.2680 \\
 & 1.2 & 2.94286 & & 6556.77 & & 0.7130 &  0.2680 \\
 & 2.2 & 3.33212 & & 6479.03 & & 0.7130 &  0.2680 \\
 & 2.6 & 3.44310 & & 6406.05 & & 0.7130 &  0.2680 \\
\hline
M1.3-dif & 0.5 & 2.66146 & 0.26\% & 6615.92 & 0.99\% & 0.7390 & 0.2430 \\
 & 1.2 & 2.95386 & 0.37\% & 6592.50 & 0.54\% & 0.7756 & 0.2077 \\
 & 2.2 & 3.33165 & 0.014\% & 6483.36 & 0.067\% & 0.8033 & 0.1810 \\
 & 2.6 & 3.43819 & 0.14\% & 6411.80 & 0.09\% & 0.7936 & 0.1901 \\
\hline 
\hline
M1.35-hom & 0.5 & 3.18750 & & 6681.75 & & 0.7130 & 0.2680 \\
 & 1 & 3.42768 & & 6678.21 & & 0.7130 & 0.2680 \\
 & 1.8 & 3.84282 & & 6588.55 & & 0.7130 & 0.2680 \\
 & 2.2 & 3.98263 & & 6495.91 & & 0.7130 & 0.2680 \\
\hline
M1.35-dif & 0.5 & 3.19352 & 0.19\% & 6726.98 & 0.68\% & 0.7637 & 0.2191 \\
 & 1 & 3.43558 & 0.23\% & 6698.23 & 0.30\% & 0.8075 & 0.1771 \\
 & 1.8 & 3.84034 & 0.06\% & 6570.07 & 0.28\% & 0.8429 & 0.1428 \\
 & 2.2 & 3.97924 & 0.08\% & 6481.87 & 0.22\% & 0.8214 & 0.1631 \\
\hline \hline
M1.45-hom & 1 & 4.82217 & & 6906.53 & & 0.7130 & 0.2680 \\
 & 1.2 & 4.98999 & & 6842.74 & & 0.7130 & 0.2680 \\
 & 1.6 & 5.23299 & & 6709.96 & & 0.7130 & 0.2680 \\
 & 1.8 & 5.33777 & & 6607.09 & & 0.7130 & 0.2680\\
\hline
M1.45-dif & 1 & 5.13913 & 6.57\% & 6769.24 & 1.99\% & 0.9948 & 0.0010 \\
 & 1.2 & 5.32206 & 6.65\% & 6676.52 & 2.43\% & 0.9943 & 0.0014 \\
 & 1.6 & 5.58805 & 6.35\% & 6469.64 & 3.71\% & 0.9823 & 0.0104 \\
 & 1.8 & 5.71755 & 7.11\% & 6347.46 & 3.93\% & 0.9660 & 0.0250 \\
\hline
\end{tabular}
\end{center}
\end{table*}

\section{Tests of internal structure}

The oscillation frequencies of our models have been computed with an updated version of the adiabatic code described in \citet{brassard92}, for values of the azimuthal degree $l$ between 0 and 3 and different values of the radial order $n$. We then studied in detail different combinations of these frequencies which could lead to observational tests of the internal structure and chemical composition of the stars.

The ``large separations'' represent the frequency differences between two modes of the same degree $l$ and successive radial orders $n$:
\begin{equation}
\Delta \nu_{n,l} = \nu_{n,l} - \nu_{n-1,l}
\end {equation}
For p-modes in the asymptotic theory ($n \gg l$), the large separations are nearly constant. The so-called ``echelle diagrams'' present the frequencies in ordinates, and the same frequencies modulo the average large separation in absissae. The asymptotic theory predicts a vertical line for each degree. 

The differences between the lines $l=0$ and $l=2$ on the one hand, $l=1$ and $l=3$ on the other hand, for each case, are the so-called ``small separations'', which we normalize as suggested by \citet{RV01} :
\begin{equation}
D\nu_{l,l+2} = \frac{1}{2l+3}(\nu_{n,l} - \nu_{n-1,l+2})
\end{equation}
The deviations of the small separations to zero show departure from the asymptotic theory. The small separations are mainly sensitive to the stellar core.

\begin{figure}[h!]
\begin{center}
\includegraphics[angle=0,height=3.5cm,width=4.3cm]{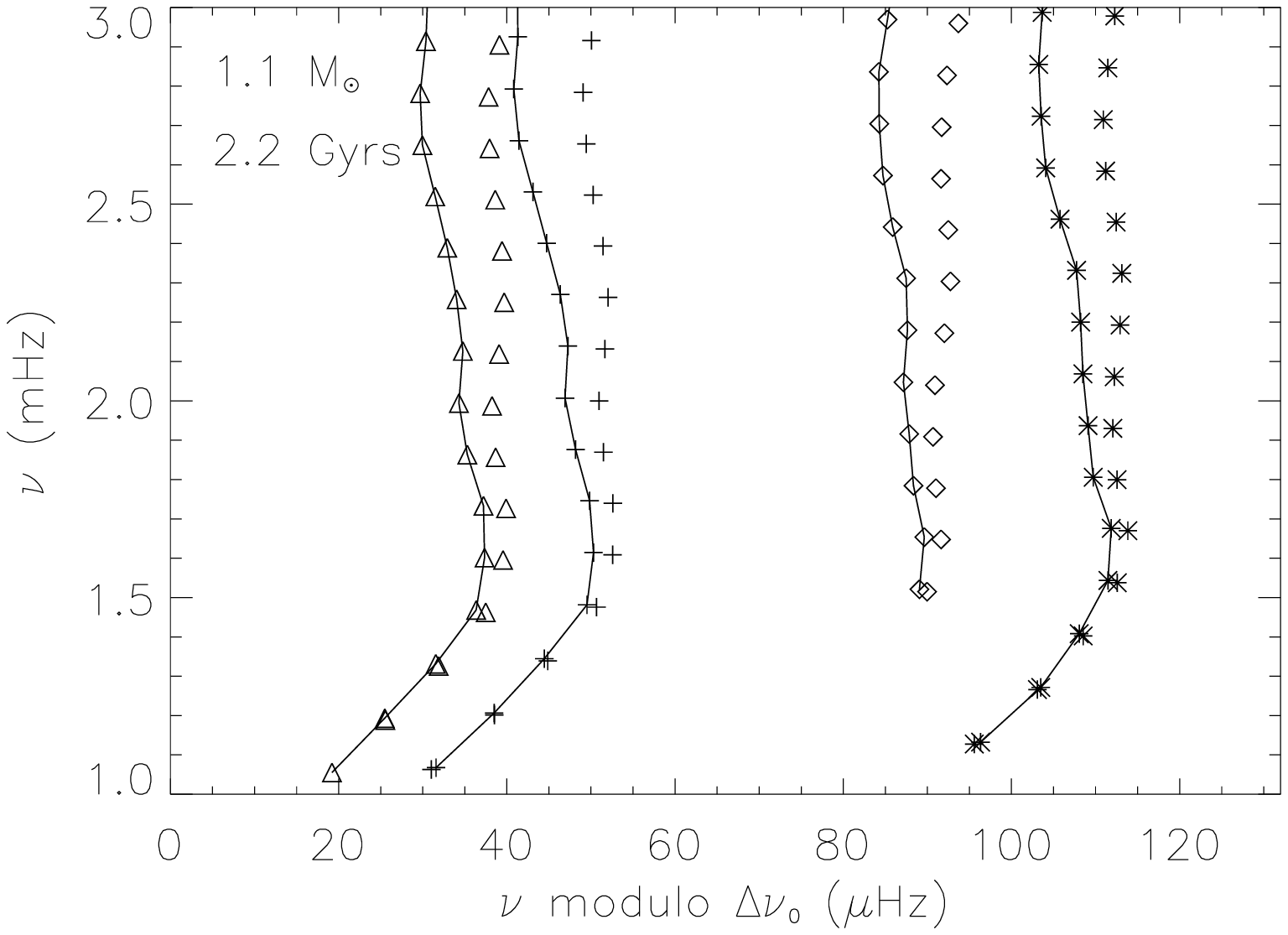} \includegraphics[angle=0,height=3.5cm,width=4.3cm]{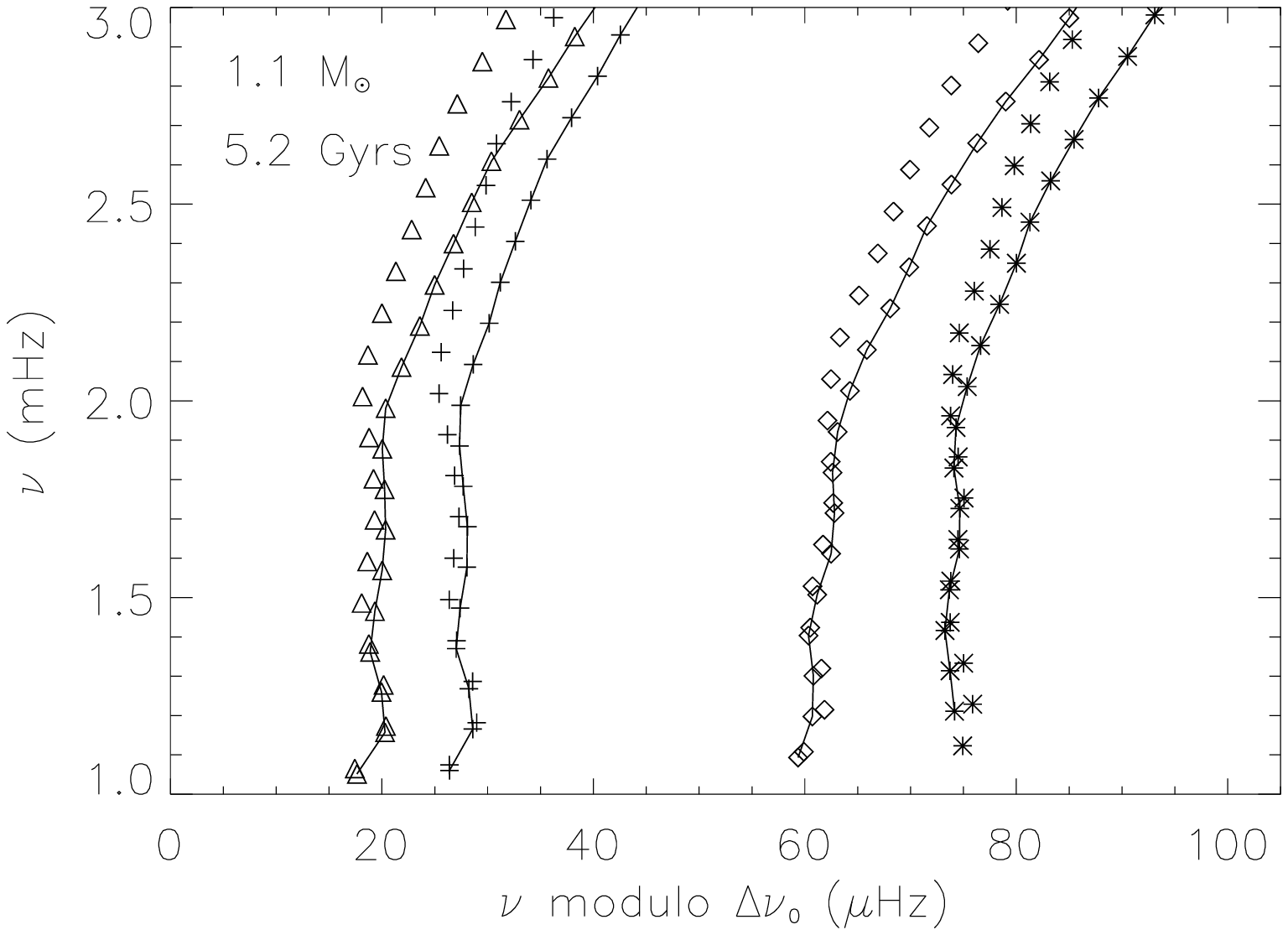} \\
\includegraphics[angle=0,height=3.5cm,width=4.3cm]{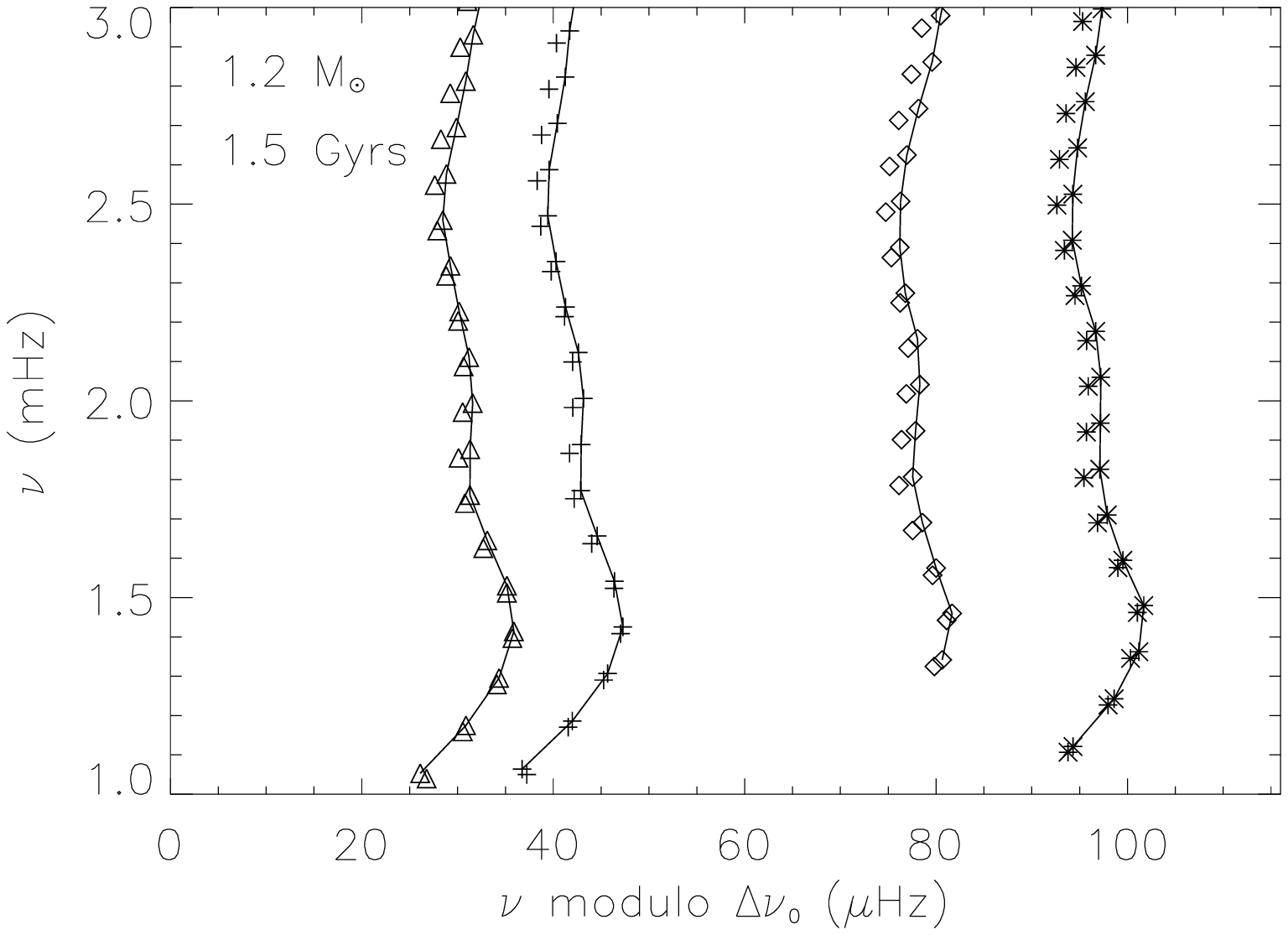} \includegraphics[angle=0,height=3.5cm,width=4.3cm]{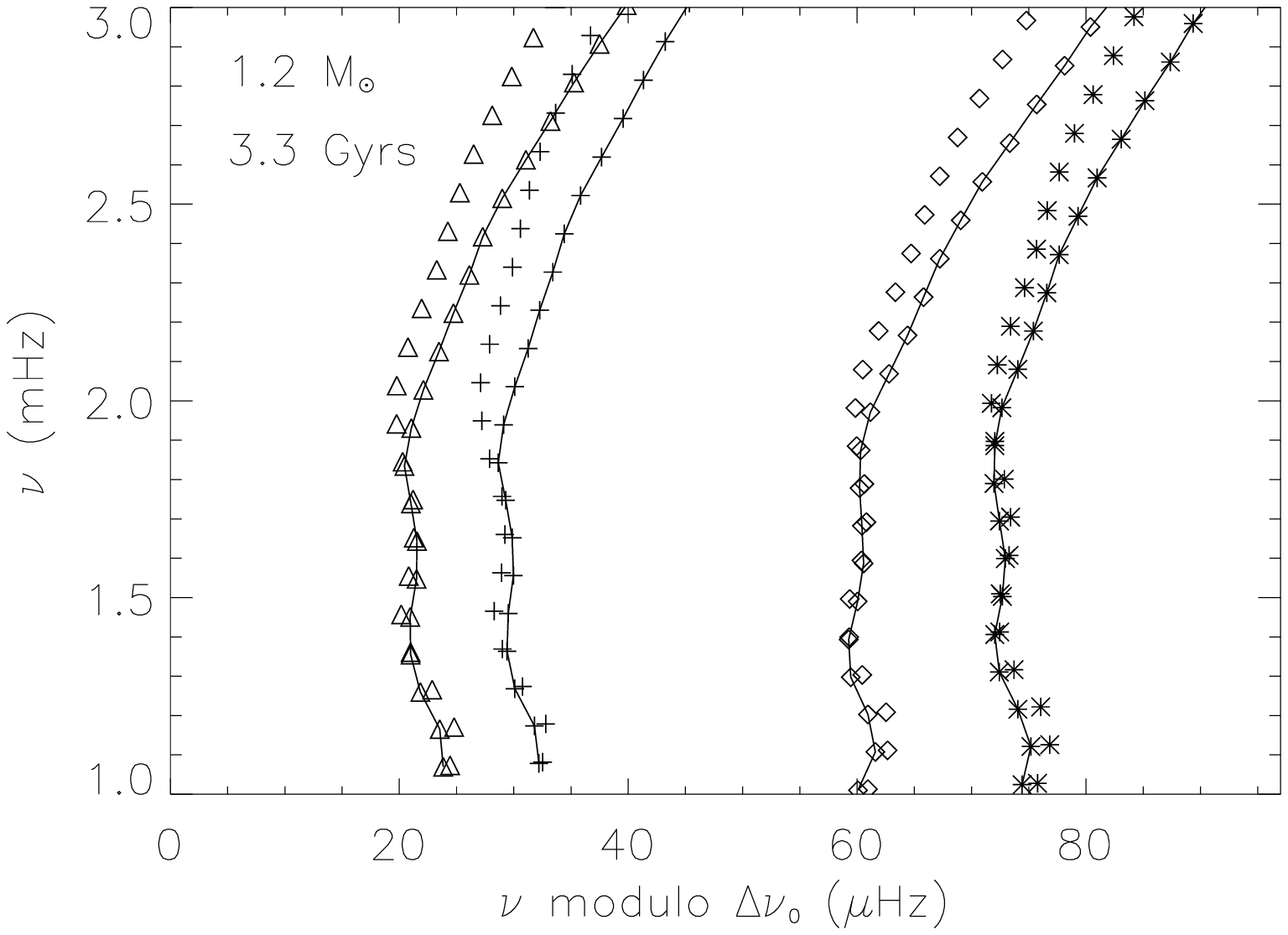} \\
\includegraphics[angle=0,height=3.5cm,width=4.3cm]{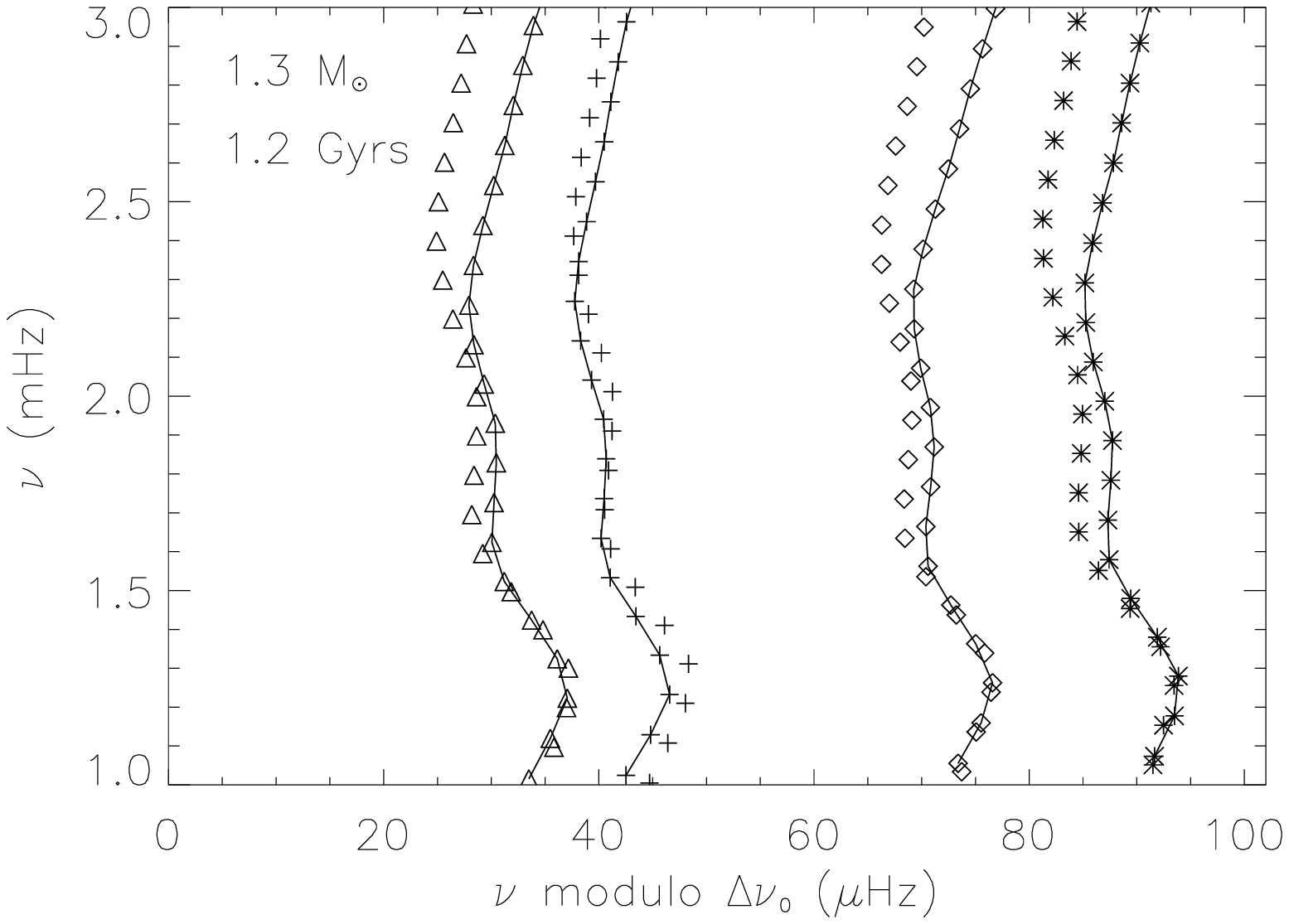} \includegraphics[angle=0,height=3.5cm,width=4.3cm]{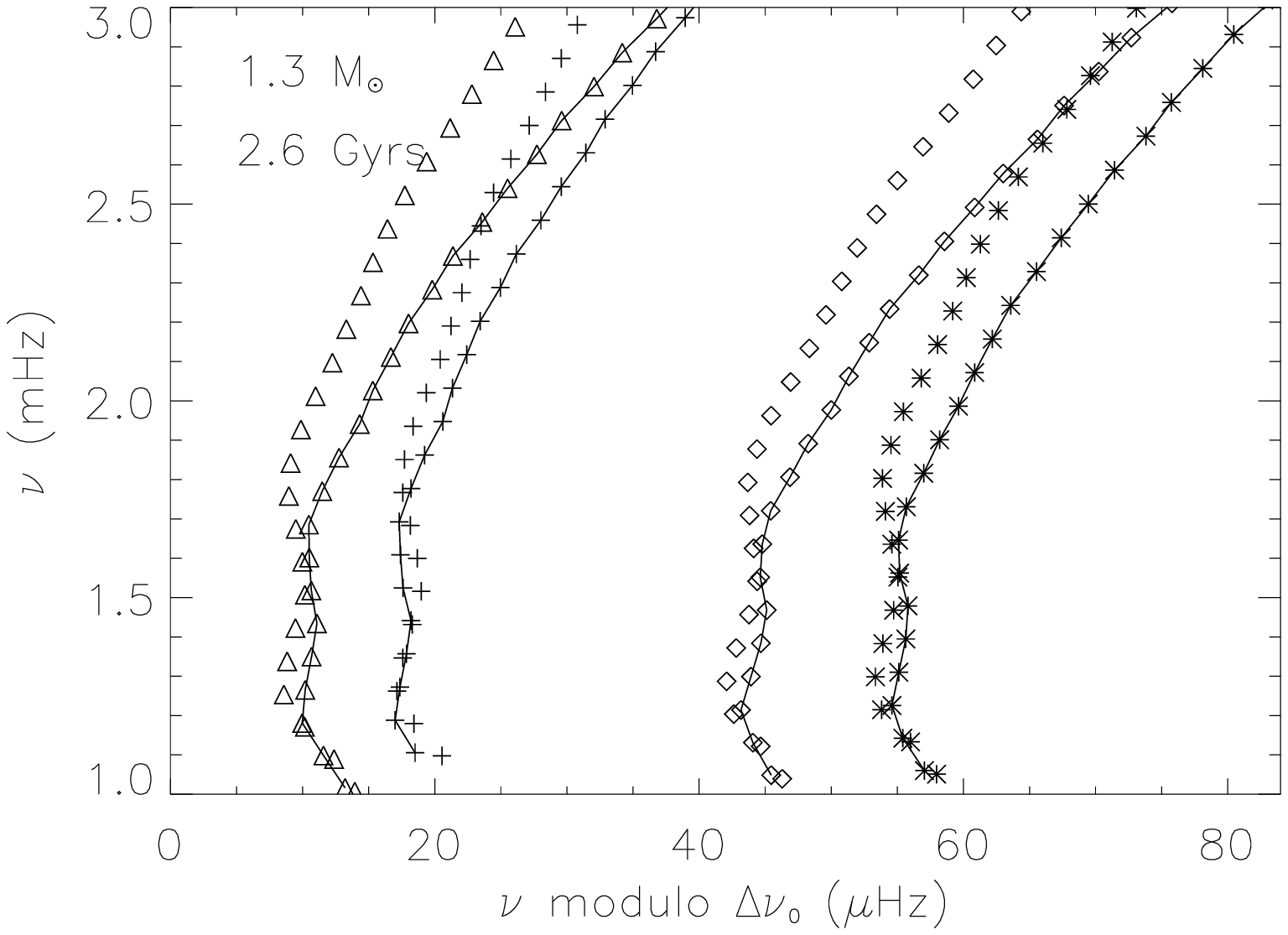} \\
\includegraphics[angle=0,height=3.5cm,width=4.3cm]{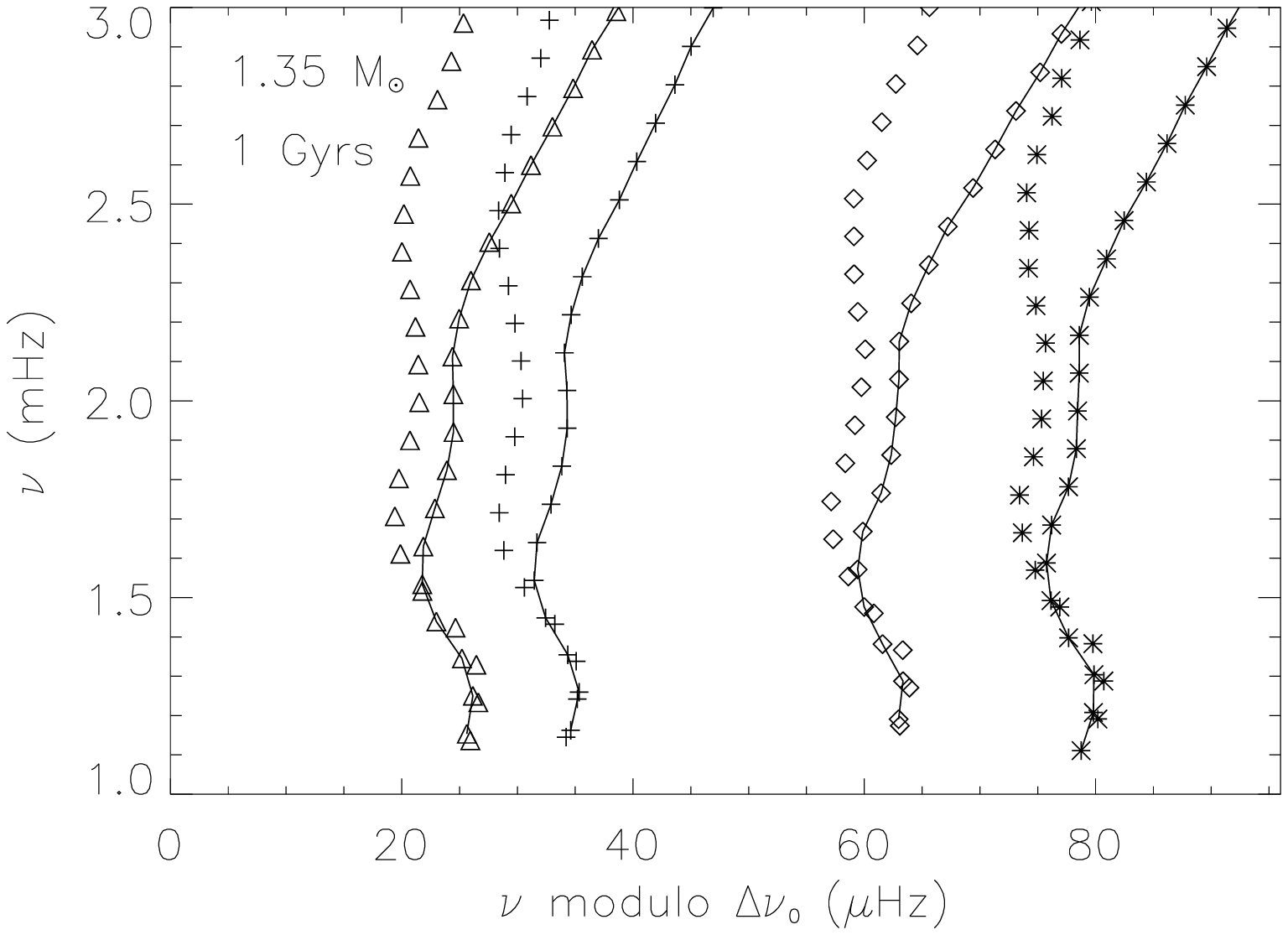} \includegraphics[angle=0,height=3.5cm,width=4.3cm]{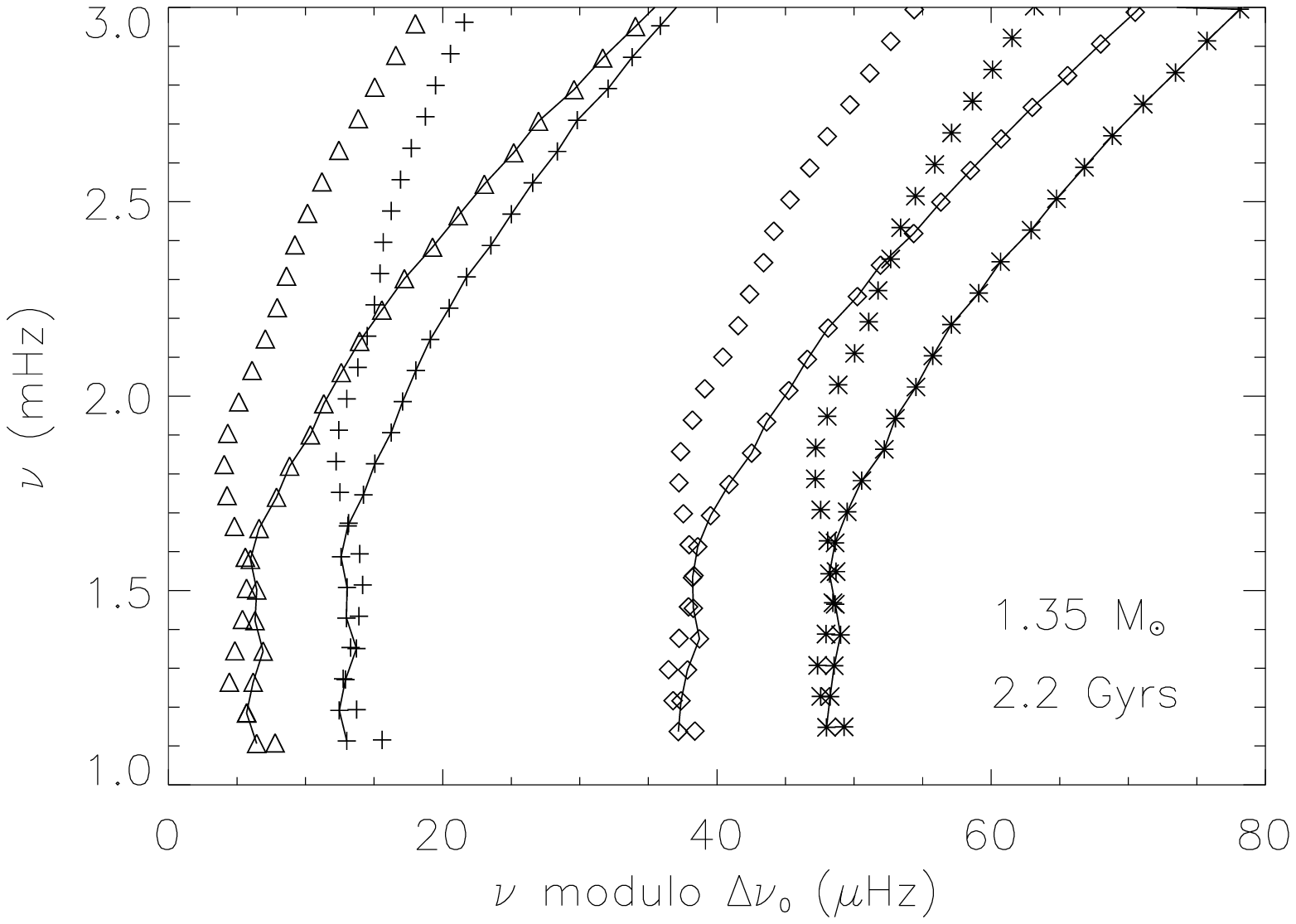} 
\end{center}
\caption{Echelle diagram for models M1.1 to M1.35 at different ages. Crosses : $l=0$, stars : $l=1$, triangles : $l=2$, diamonds : $l=3$. The points connected by lines are for models with diffusion.}
\label{lsep}
\end{figure}

\begin{figure}[h!]
\begin{center}
\includegraphics[angle=0,height=3.5cm,width=4.3cm]{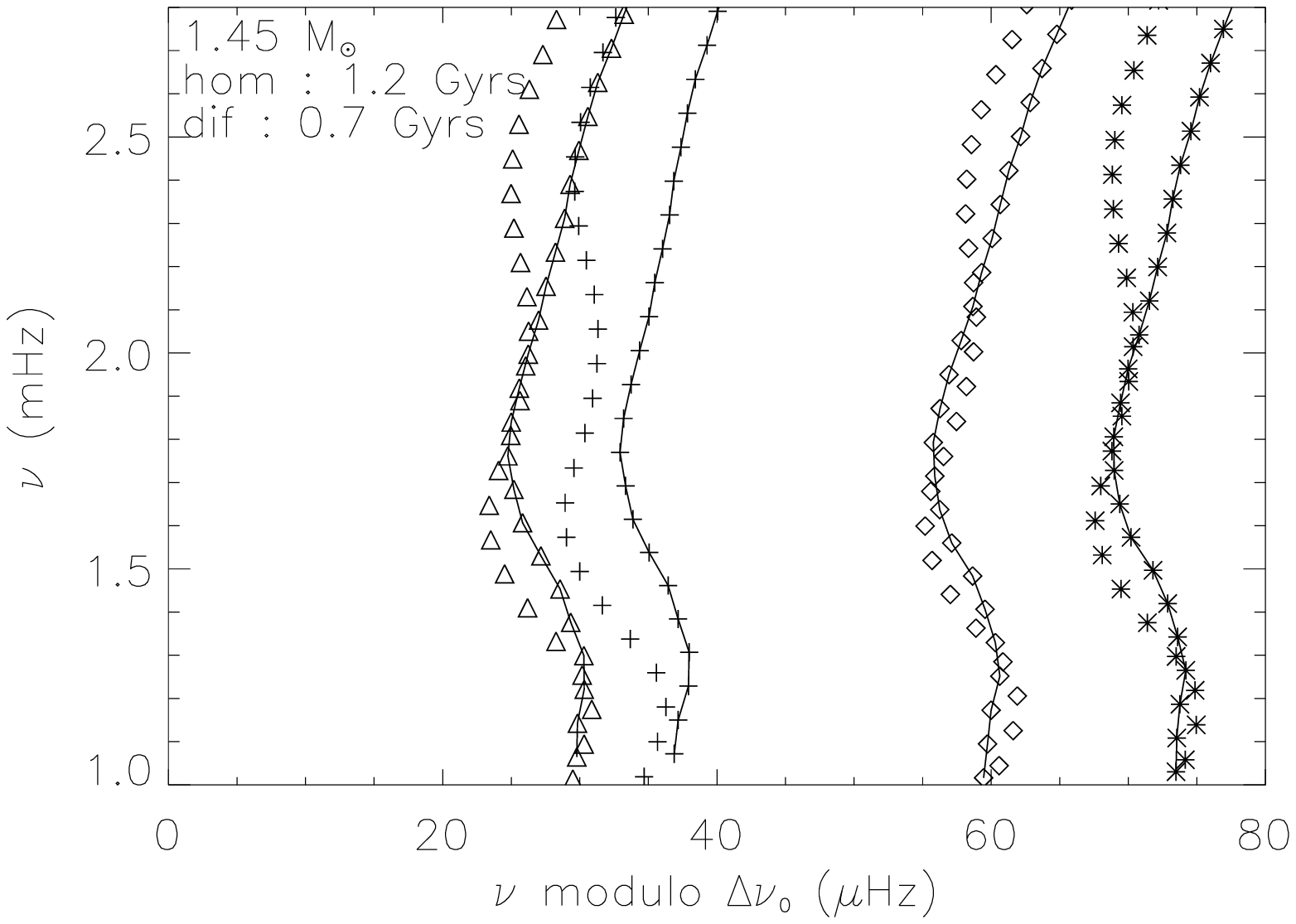} \includegraphics[angle=0,height=3.5cm,width=4.3cm]{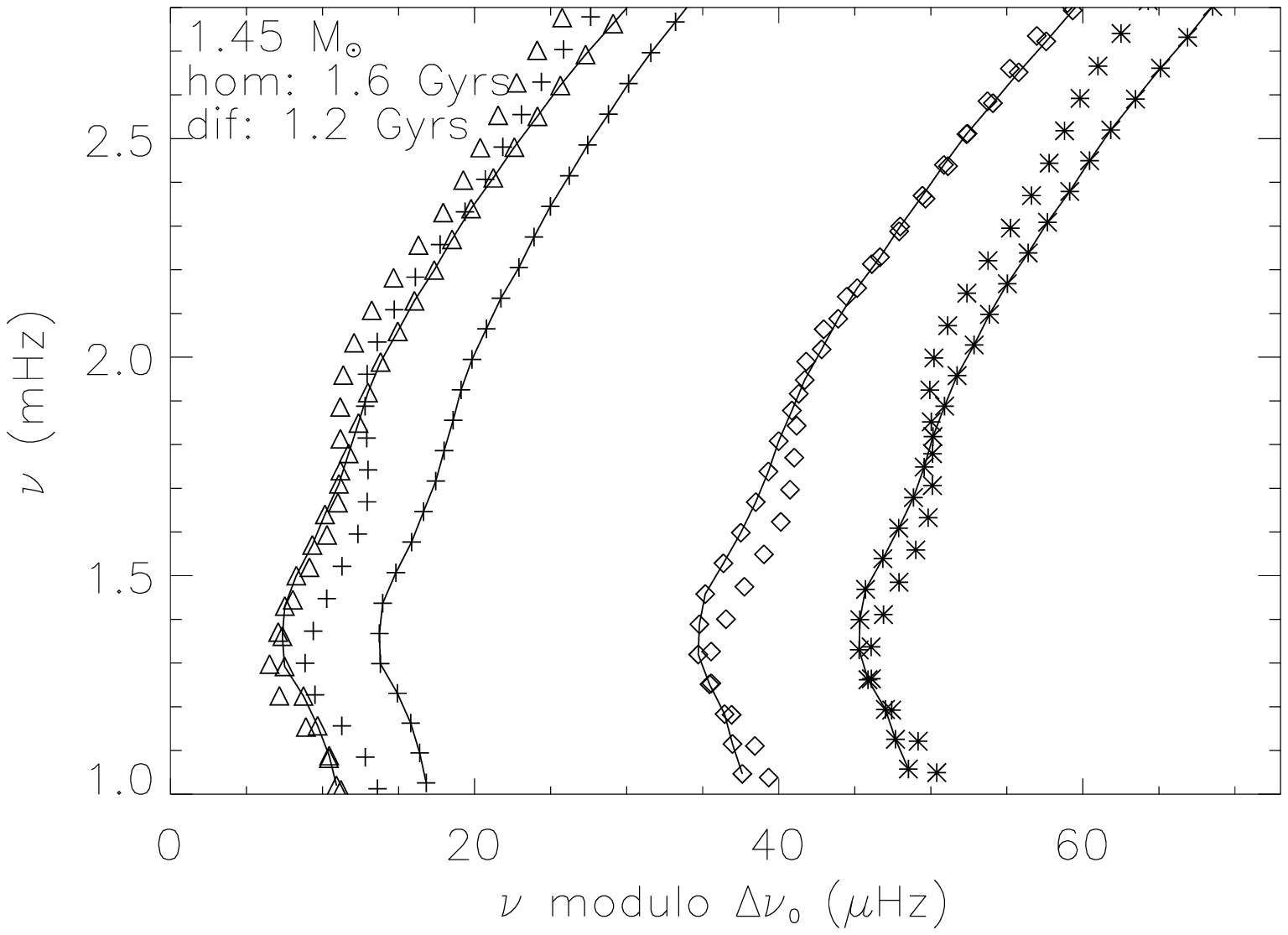}
\end{center}
\caption{Echelle diagram for models M1.45 at 1.2 Gyrs (with diffusion) and 1.6 Gyrs (without diffusion) on the left, and at 0.7 Gyrs (with diffusion) and 1.2 Gyrs (without diffusion) on the right. Crosses : $l=0$, stars : $l=1$, triangles : $l=2$, diamonds : $l=3$. The points connected by lines are for models with diffusion.}
\label{lsep145}
\end{figure}

\begin{figure}[h!]
\begin{center}
\includegraphics[angle=0,height=3.5cm,width=4.3cm]{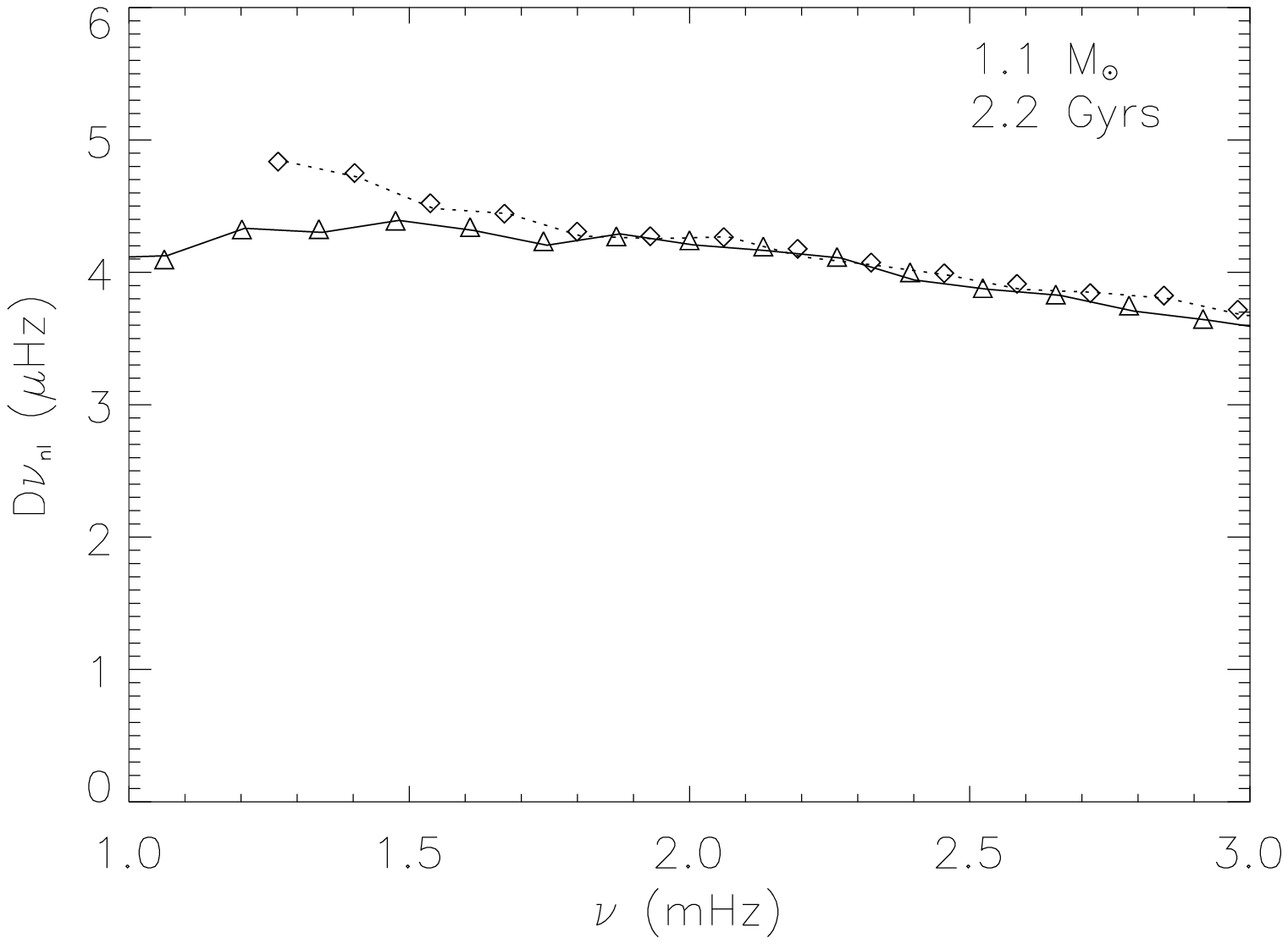} \includegraphics[angle=0,height=3.5cm,width=4.3cm]{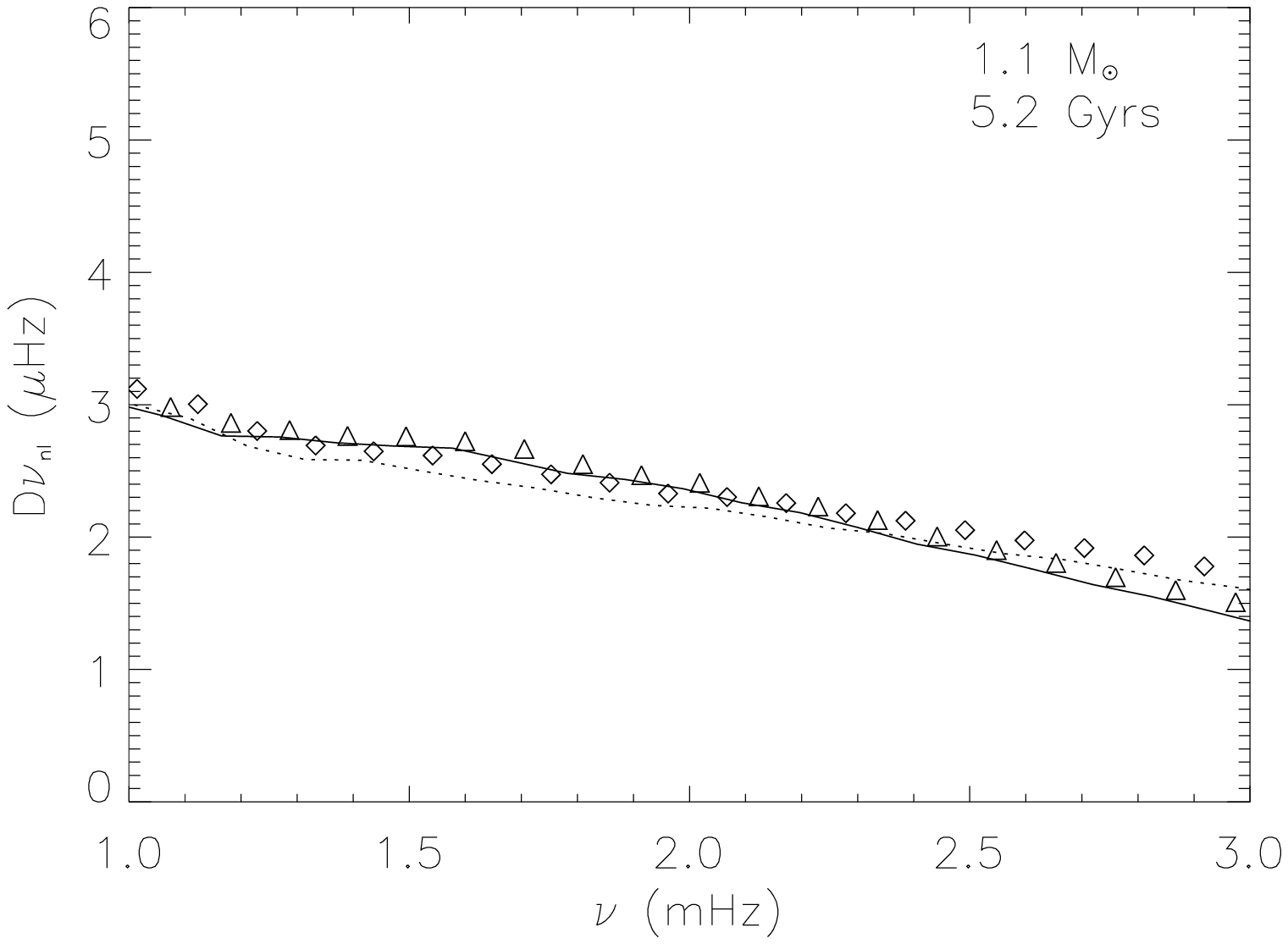} \\
\includegraphics[angle=0,height=3.5cm,width=4.3cm]{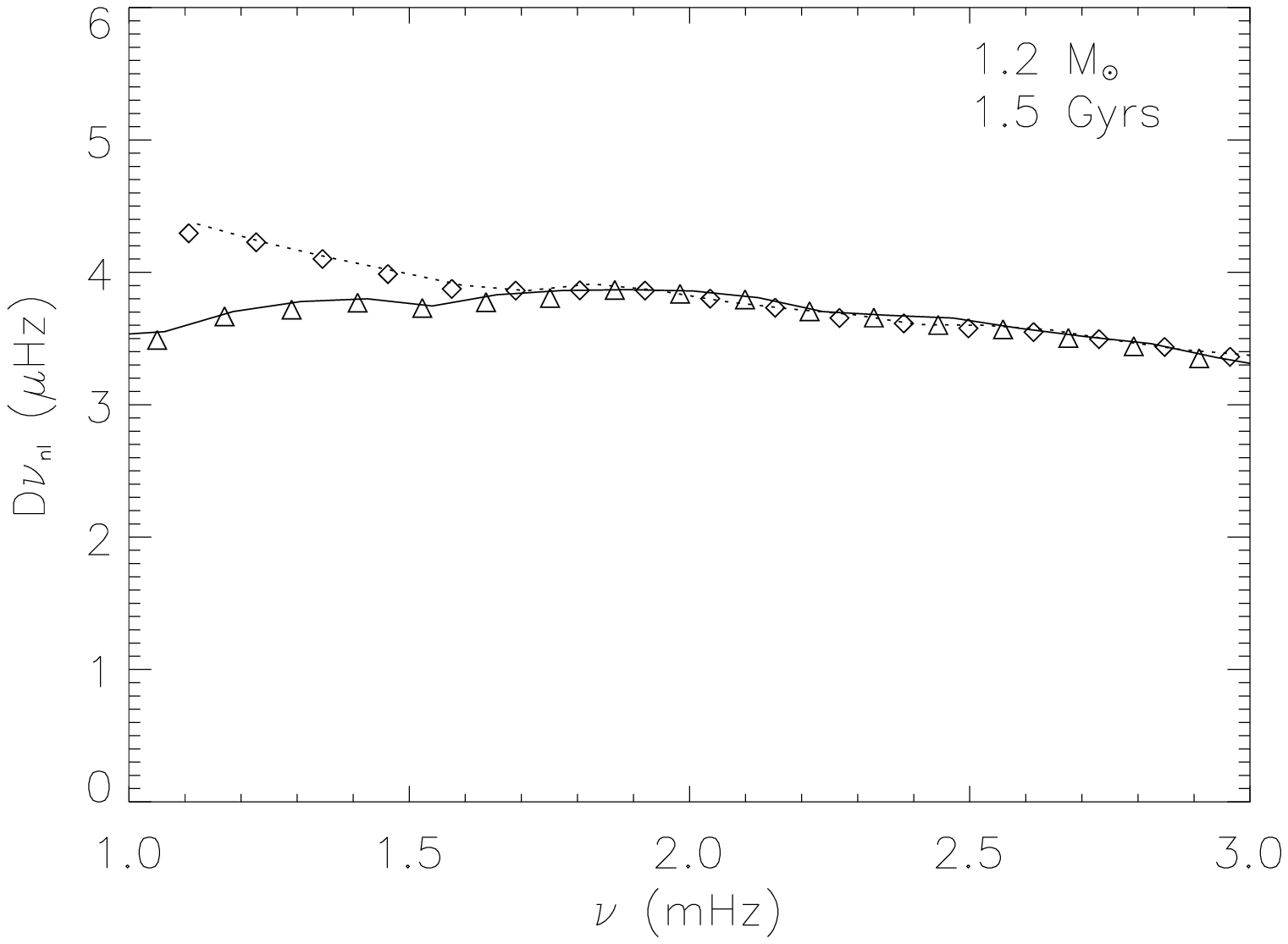} \includegraphics[angle=0,height=3.5cm,width=4.3cm]{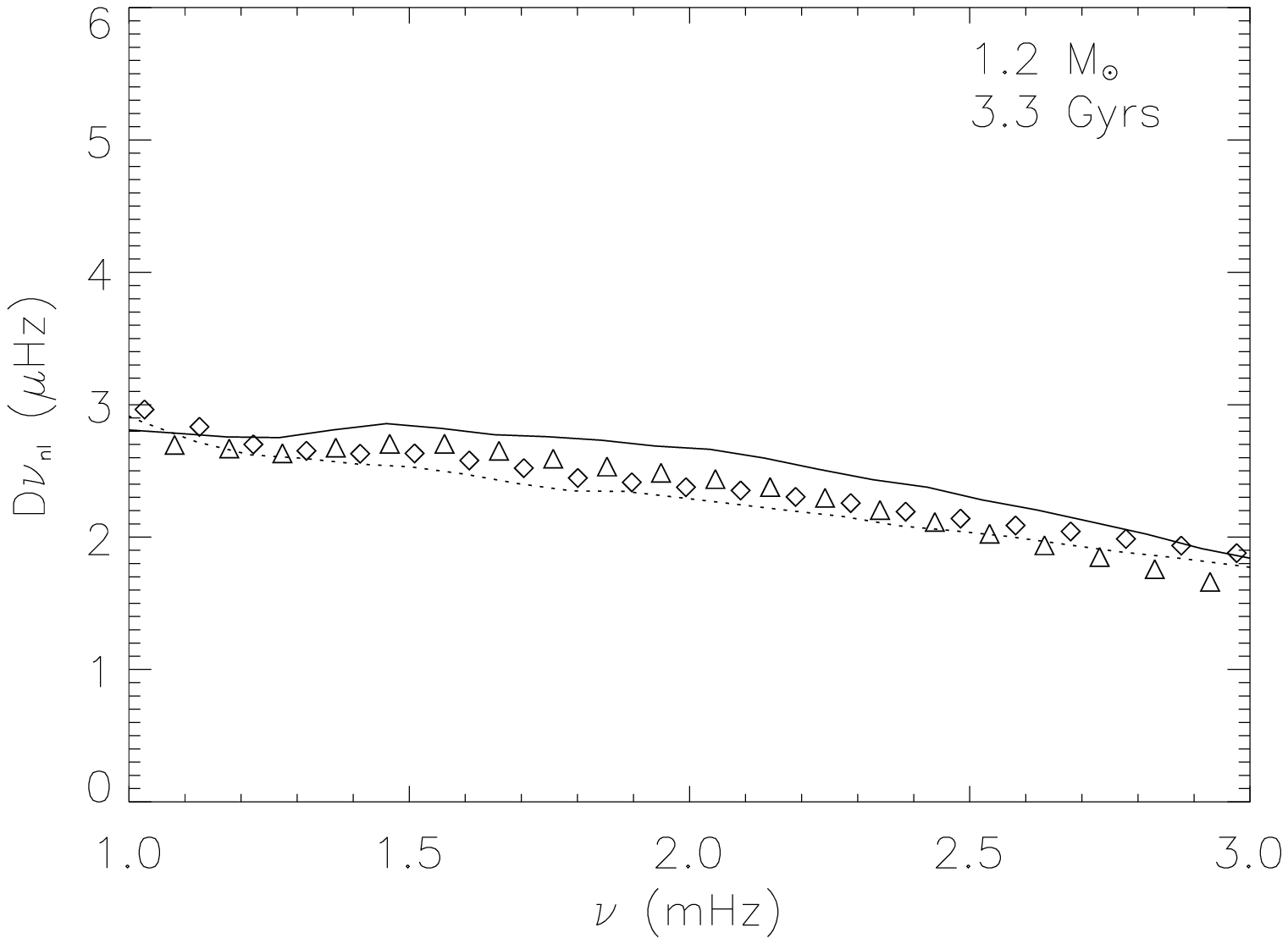} \\
\includegraphics[angle=0,height=3.5cm,width=4.3cm]{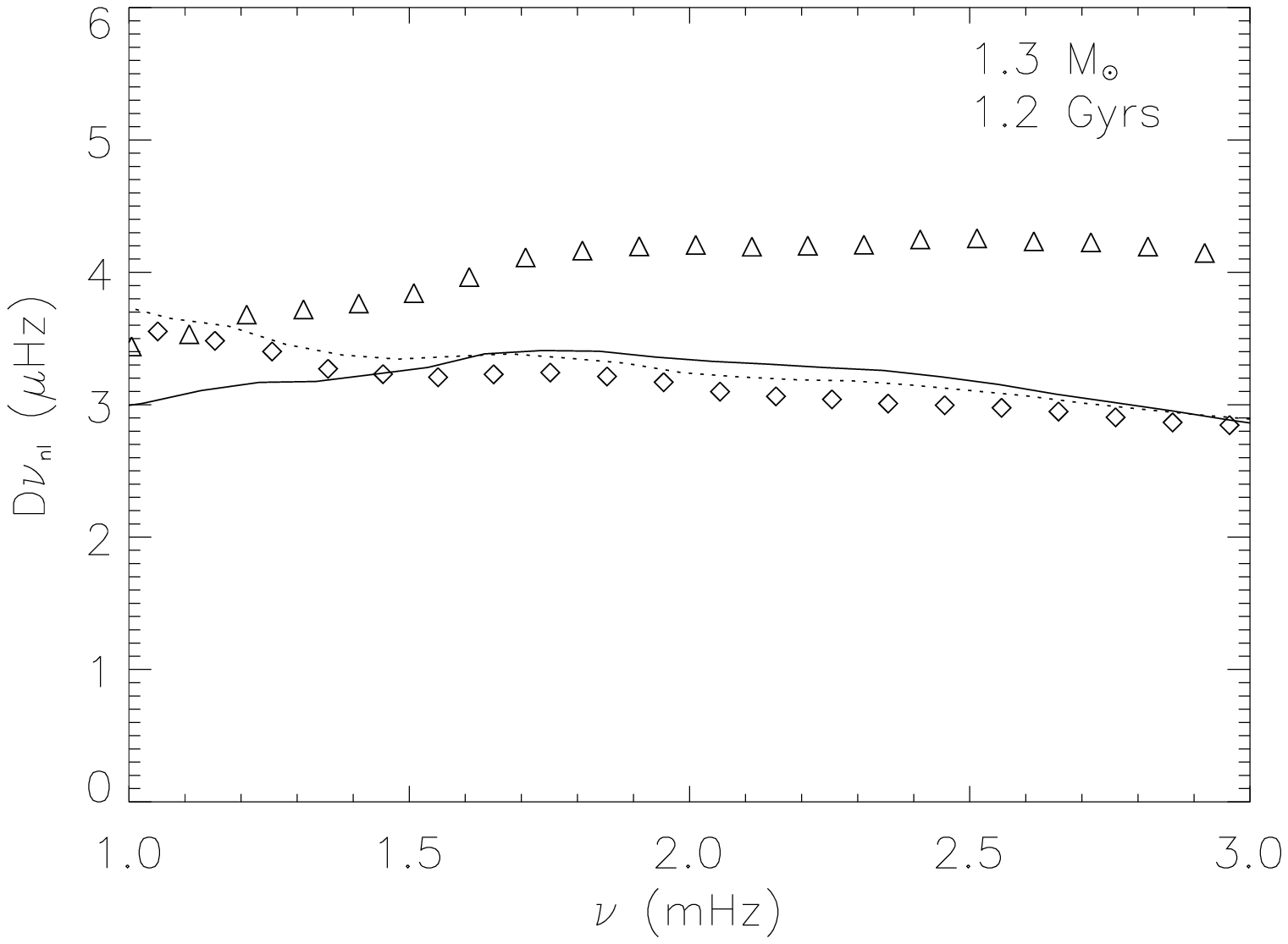} \includegraphics[angle=0,height=3.5cm,width=4.3cm]{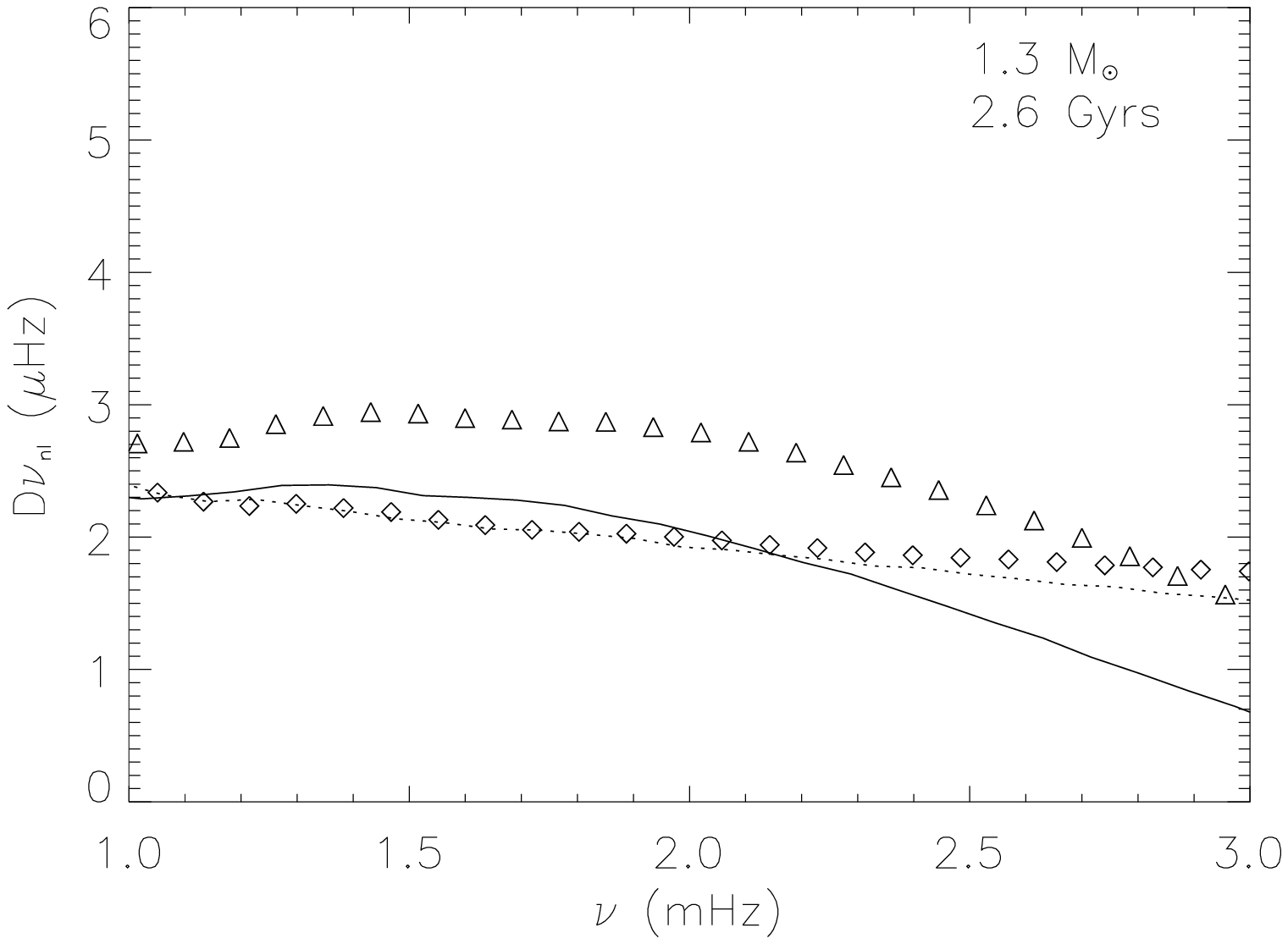} \\
\includegraphics[angle=0,height=3.5cm,width=4.3cm]{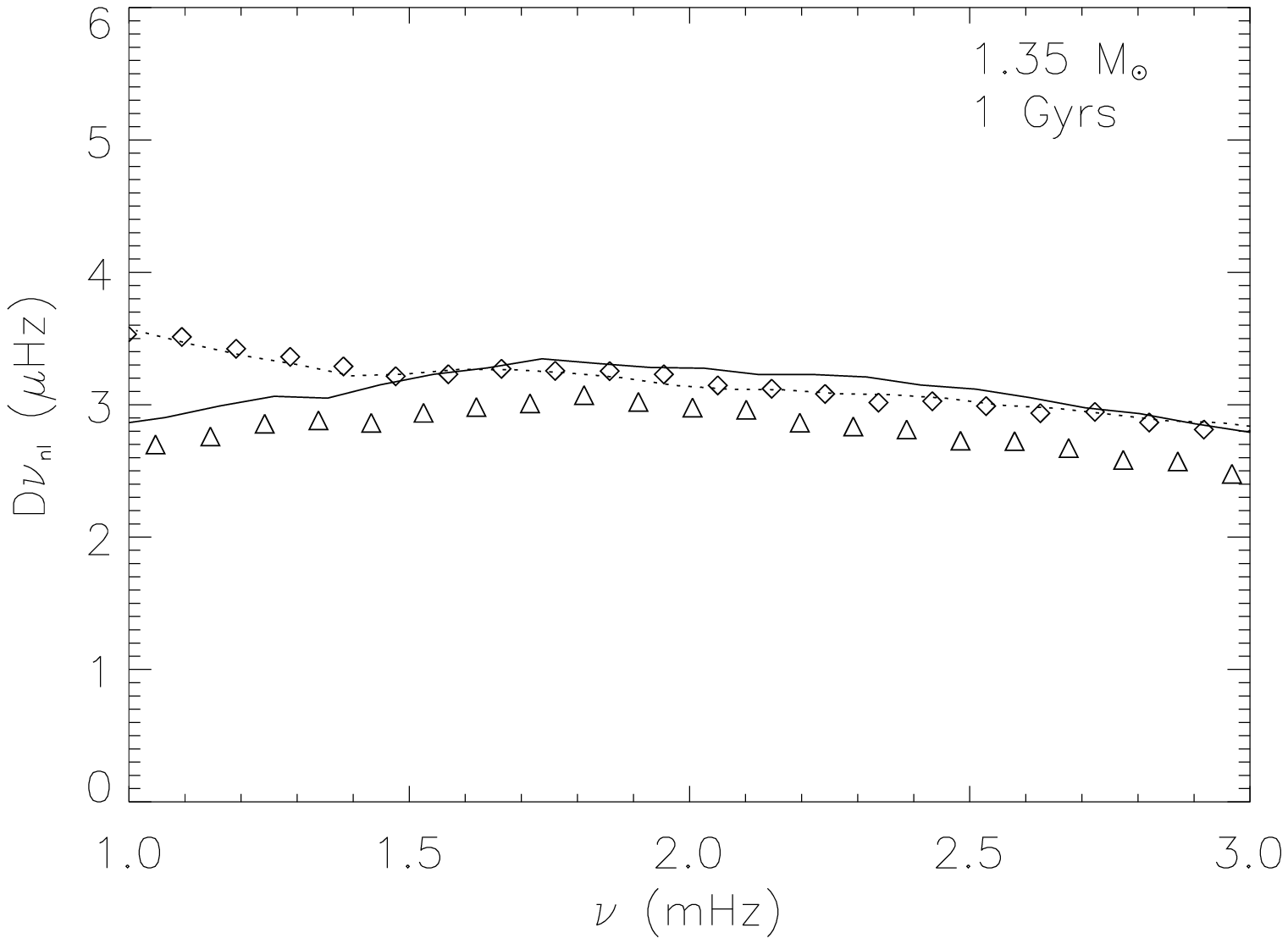} \includegraphics[angle=0,height=3.5cm,width=4.3cm]{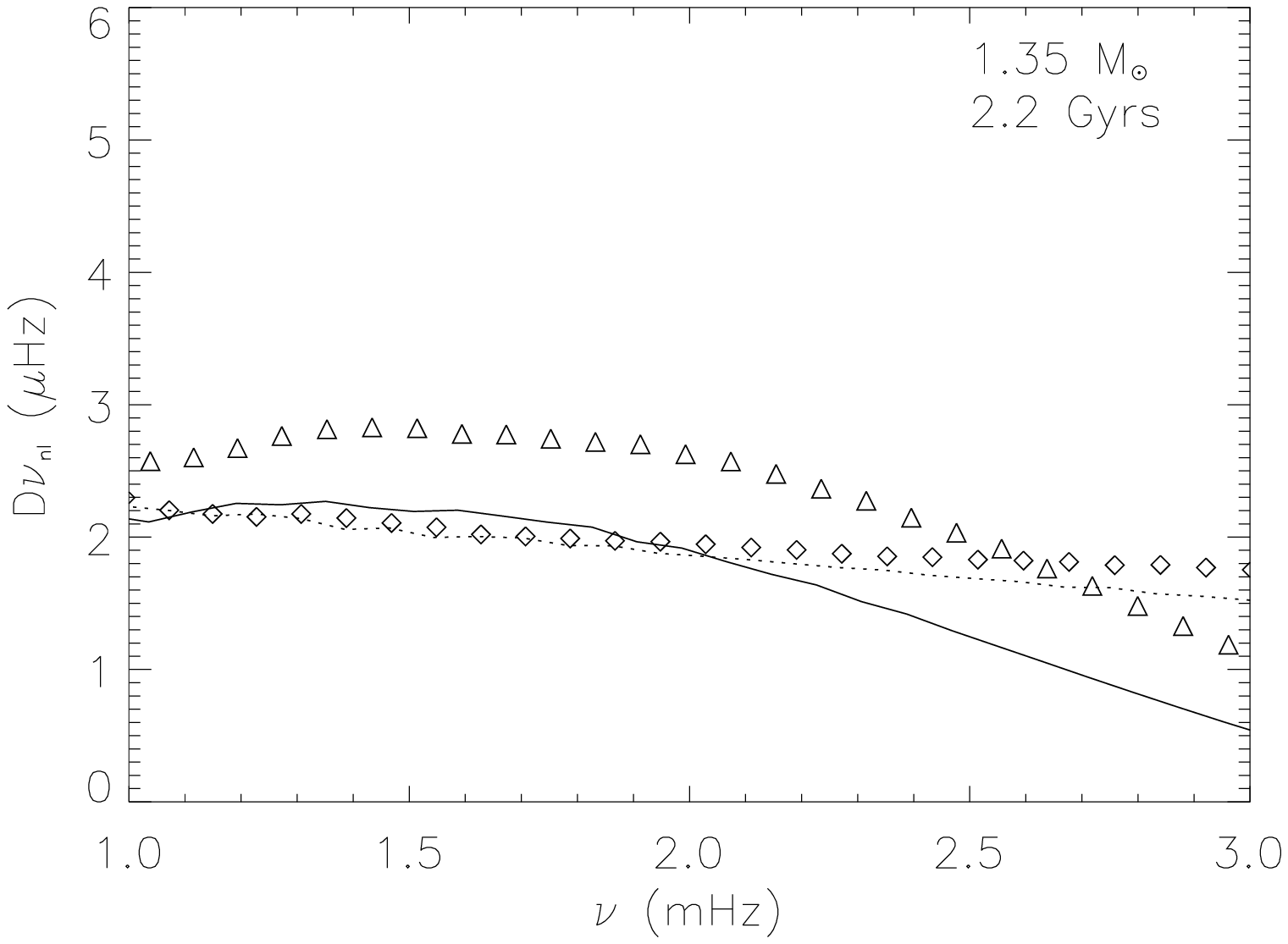} 
\end{center}
\caption{Small separations for models M1.1 to M1.35 at different ages. The lines are for models with diffusion (solid lines : $D\nu_{0,2}$, dotted lines : $D\nu_{1,3}$) and the symbols are for models without diffusion (triangles : $D\nu_{0,2}$, diamonds : $D\nu_{1,3}$).}
\label{ssep}
\end{figure}

\begin{figure}[h!]
\begin{center}
\includegraphics[angle=0,height=3.5cm,width=4.3cm]{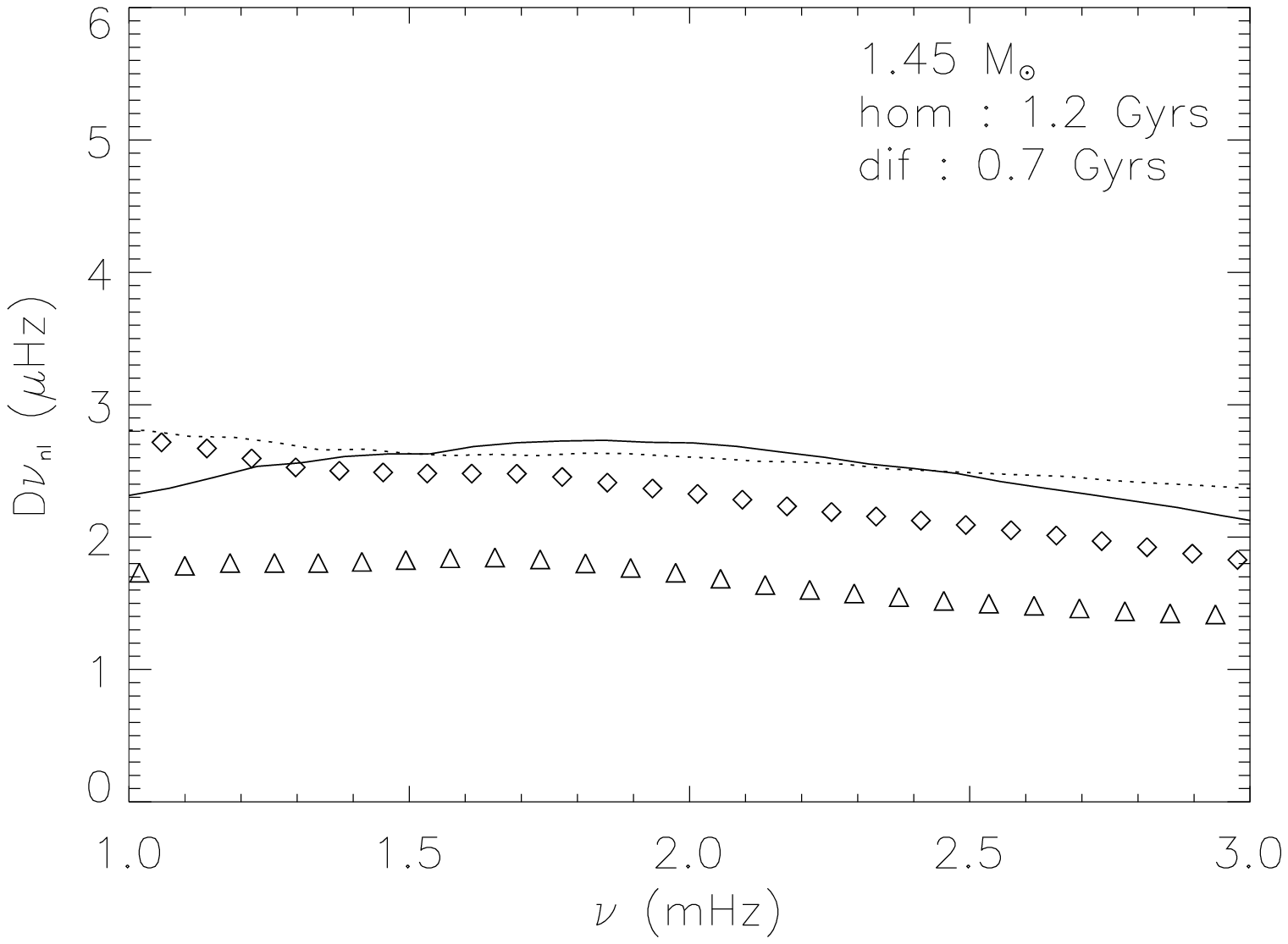} \includegraphics[angle=0,height=3.5cm,width=4.3cm]{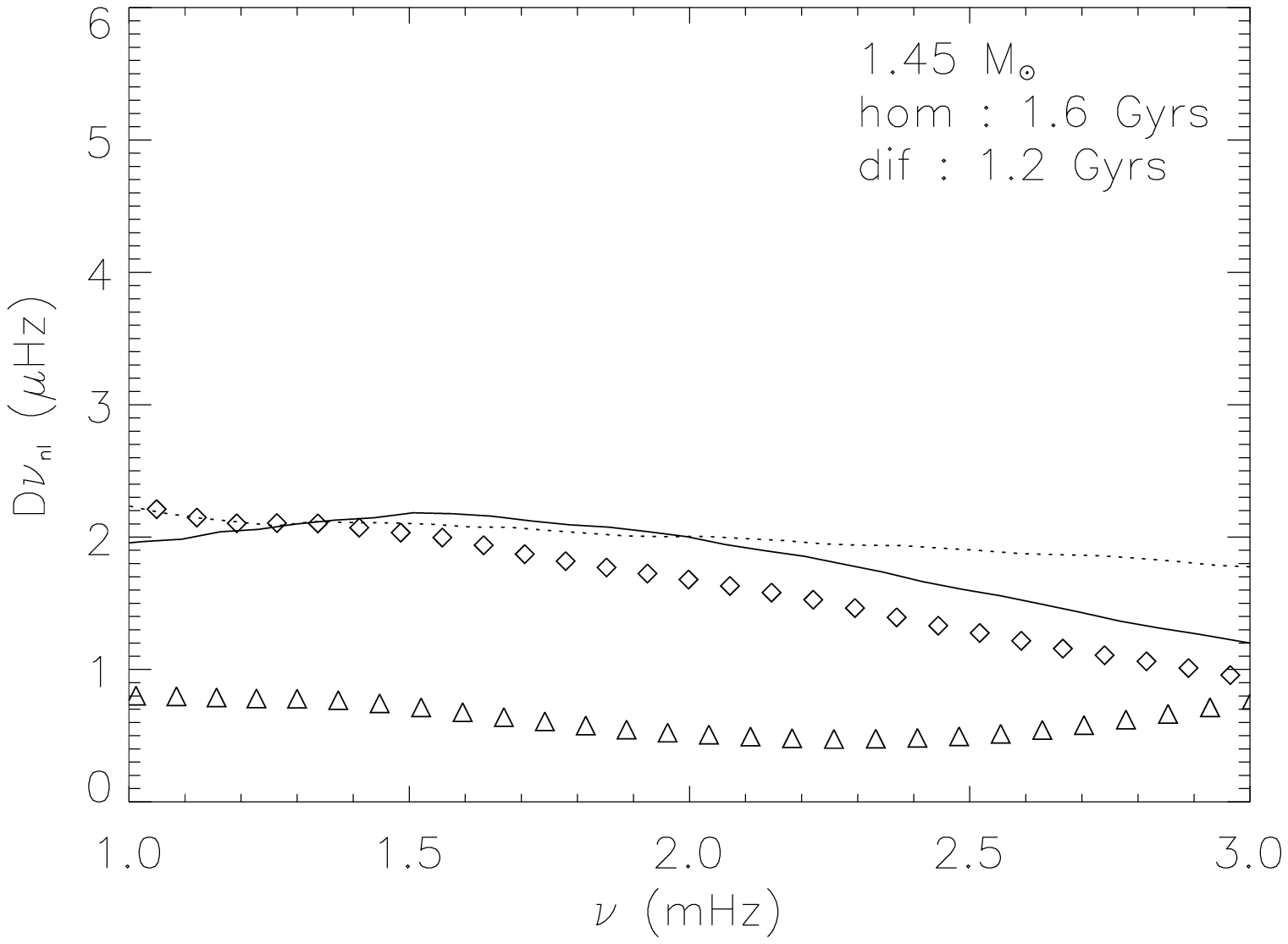}
\end{center}
\caption{Small separations for models M1.45 at 1.2 Gyrs (with diffusion) and 1.6 Gyrs (without diffusion) on the left, and at 0.7 Gyrs (with diffusion) and 1.2 Gyrs (without diffusion) on the right. The lines are for models with diffusion (solid lines : $D\nu_{0,2}$, dotted lines : $D\nu_{1,3}$) and the symbols are for models without diffusion (triangles : $D\nu_{0,2}$, diamonds : $D\nu_{1,3}$).}
\label{ssep145}
\end{figure}

In Figure \ref{lsep}, we present as an example the echelle diagram for models M1.1 to M1.35 at different ages. For M1.45, we compare echelle diagrams for models at the same location on the HR diagram, which correspond to different ages for the models with and without diffusion. This is presented in Figure \ref{lsep145}. The small separations of the same models are presented in Figure \ref{ssep} for models M1.1 to M1.35 and in Figure \ref{ssep145} for models M1.45.  

We can see on these figures that the differences between the models with and without diffusion become more important during the evolution when the mass of the star increases. Up to 1.3 \msol, the tracks in the large and small separations lie very close to each other whereas for higher masses significant deviations do appear. These effects are related to the internal differences in chemical composition and to the different extension of the convective cores.
  
According to the results of \citet{bazot05}, which presents the seismic analysis of the star $\mu$ Arae, observed with the spectrograph HARPS, the uncertainties on the small separations are about 0.37 $\mu$Hz. In the cases of 1.1 and 1.2 \msol, the differences between models with and without helium diffusion are too small to be observable. For higher masses, this difference could be detectable.

\section{Tests of helium gradients}

\subsection{The helium gradients}

\begin{figure}[h!]
\begin{center}
\includegraphics[angle=0,height=4cm,width=\columnwidth]{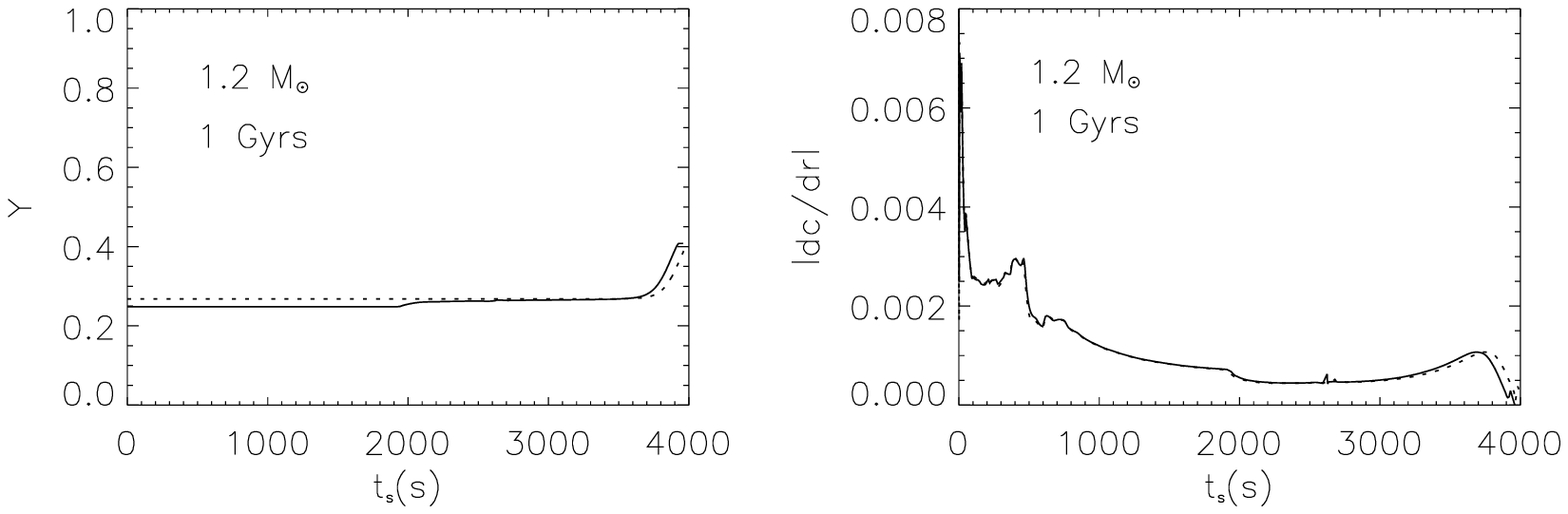}
\includegraphics[angle=0,height=4cm,width=\columnwidth]{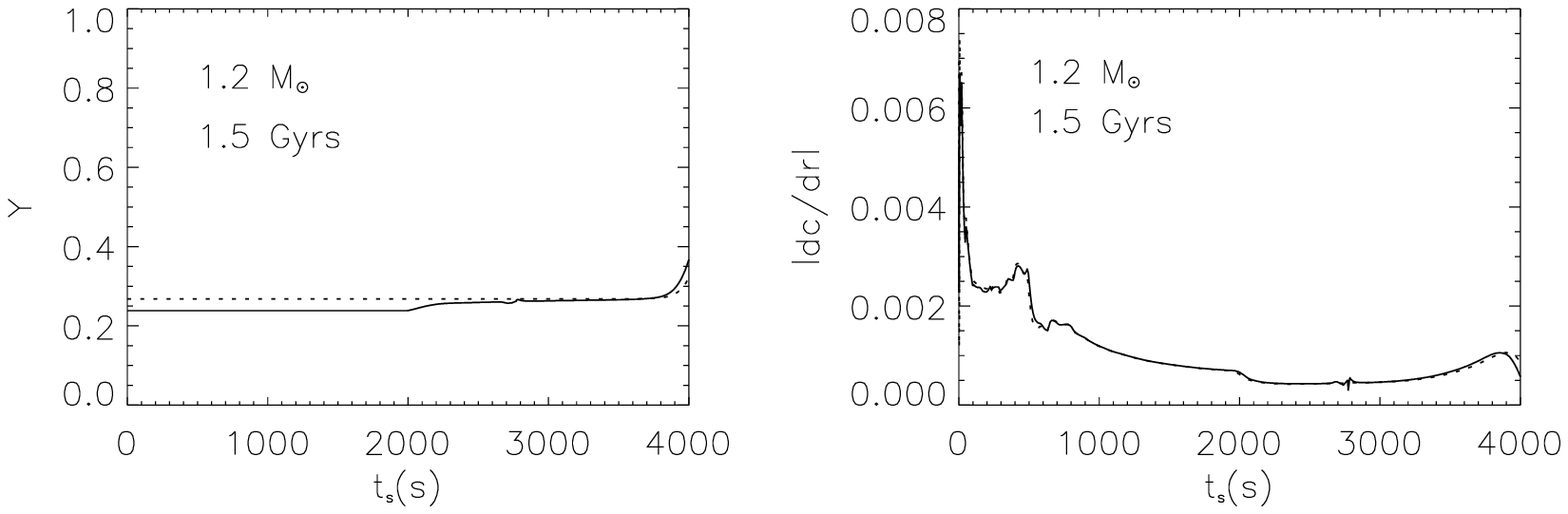}
\includegraphics[angle=0,height=4cm,width=\columnwidth]{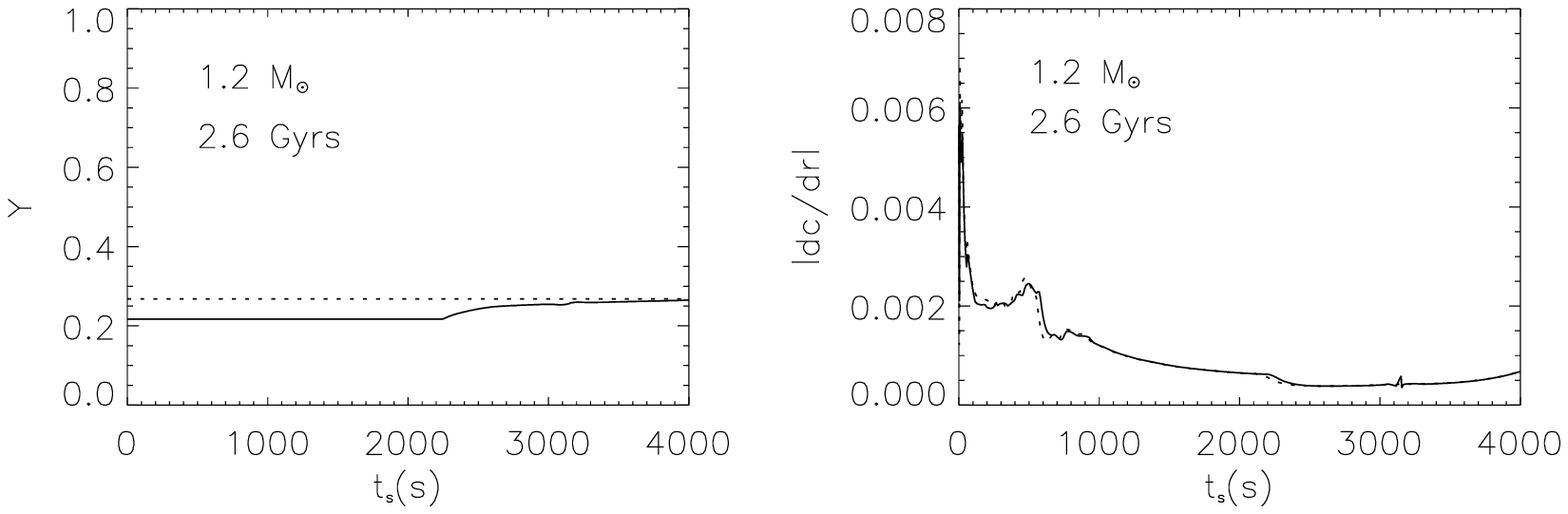}
\includegraphics[angle=0,height=4cm,width=\columnwidth]{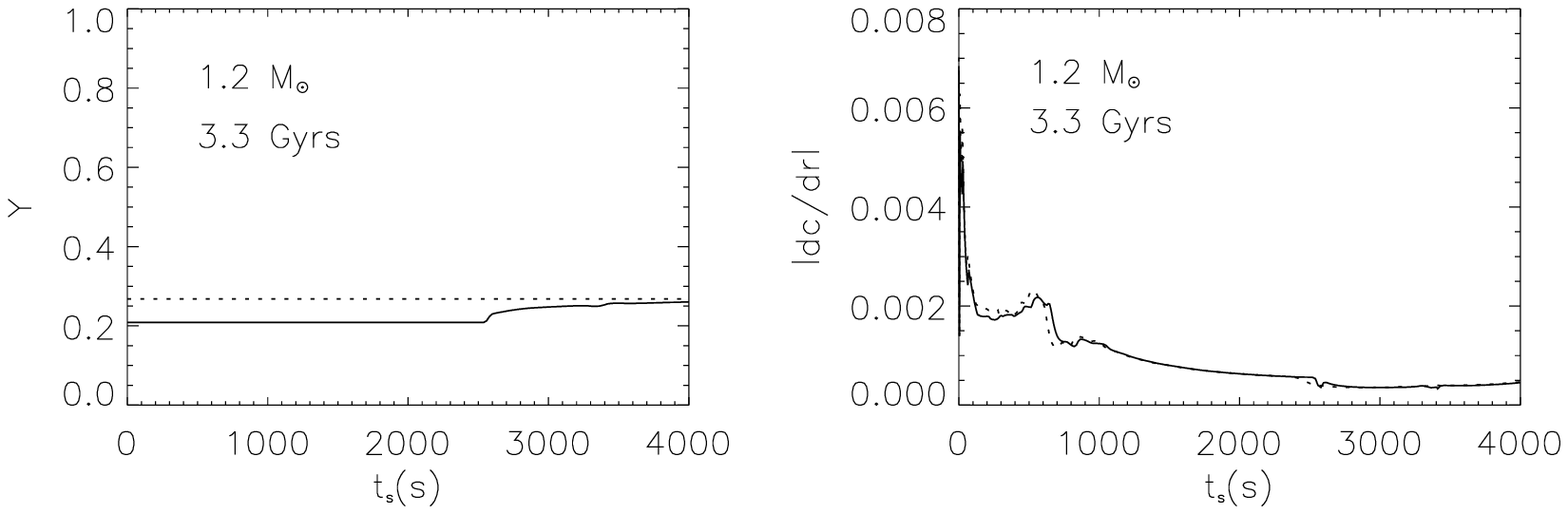}
\end{center}
\caption{Helium profiles (left column) and gradients of the sound velocity (right column) in the models with (solid lines) and without (dashed lines) element diffusion, for 1.2 \msol \ as a function of the acoustic depth at 1, 1.5, 2.6 and 3.3 Gyrs}
\label{He1.2}
\end{figure}

\begin{figure}[h!]
\begin{center}
\includegraphics[angle=0,height=4cm,width=\columnwidth]{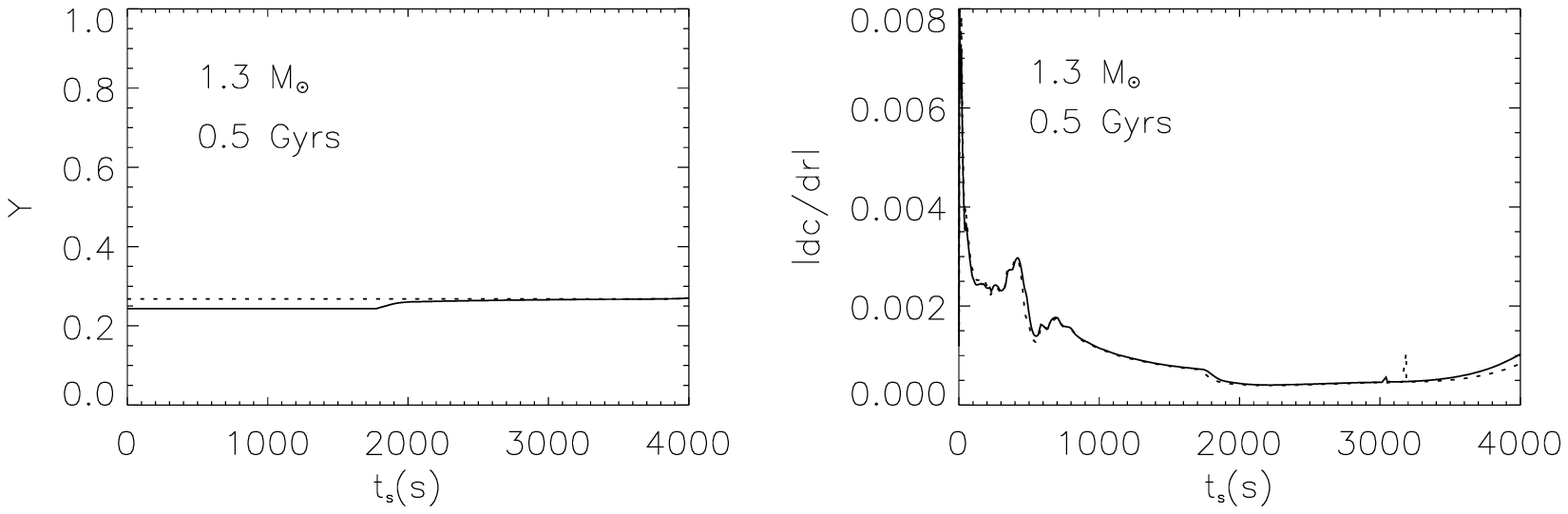}
\includegraphics[angle=0,height=4cm,width=\columnwidth]{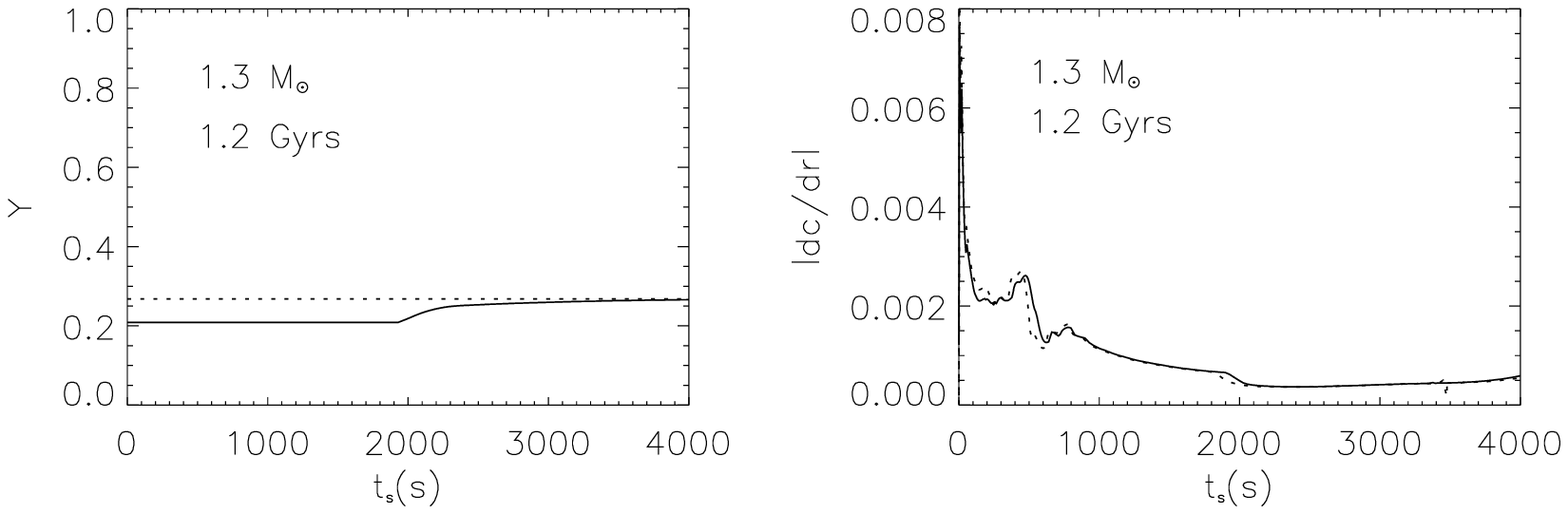}
\includegraphics[angle=0,height=4cm,width=\columnwidth]{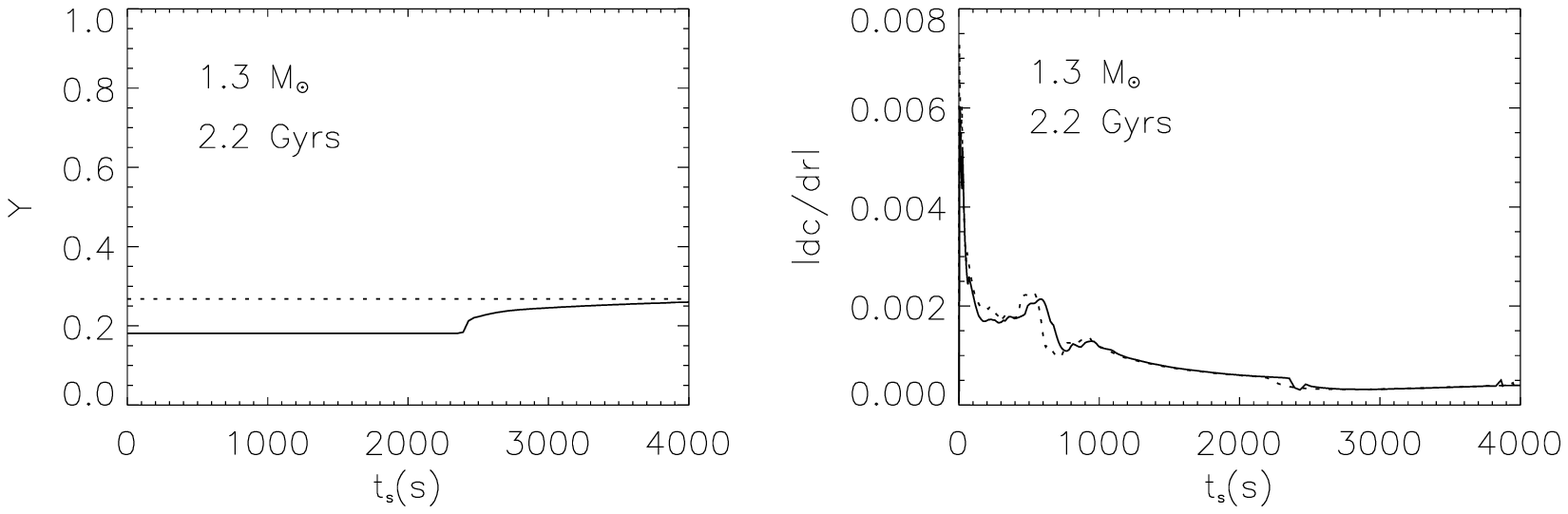}
\includegraphics[angle=0,height=4cm,width=\columnwidth]{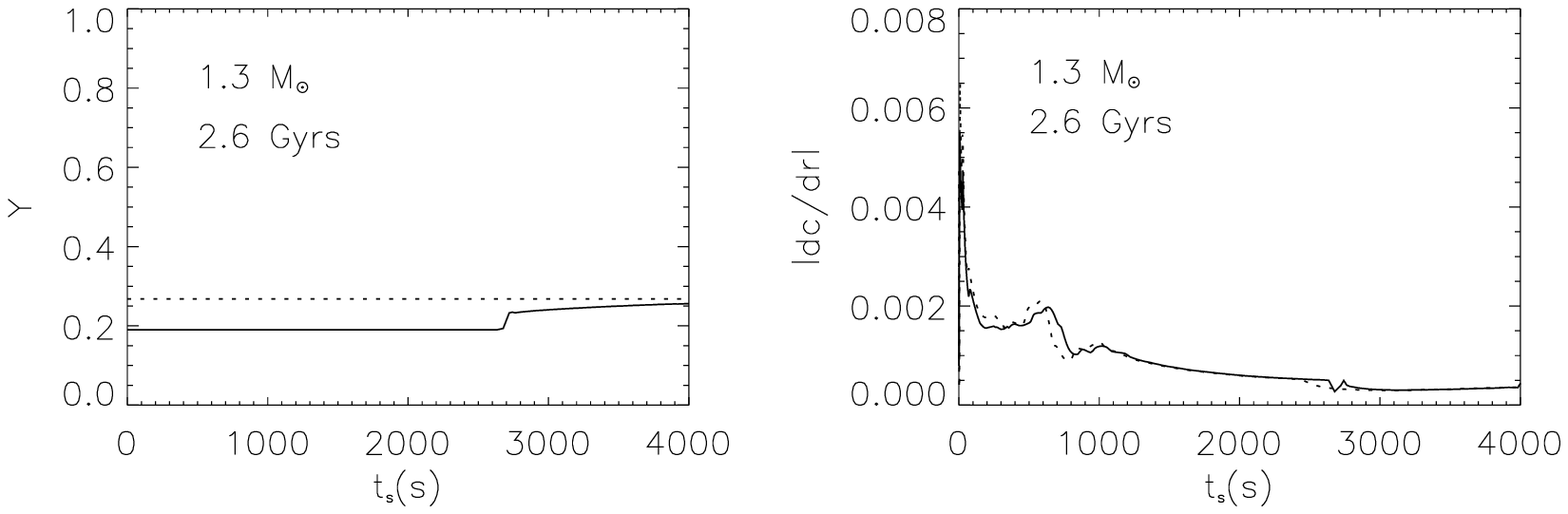}
\end{center}
\caption{Same as Fig. \ref{He1.2} for 1.3 \msol \ at 0.5, 1.2, 2.2 and 2.6 Gyrs}
\label{He1.3}
\end{figure}

\begin{figure}[h!]
\begin{center}
\includegraphics[angle=0,height=4cm,width=\columnwidth]{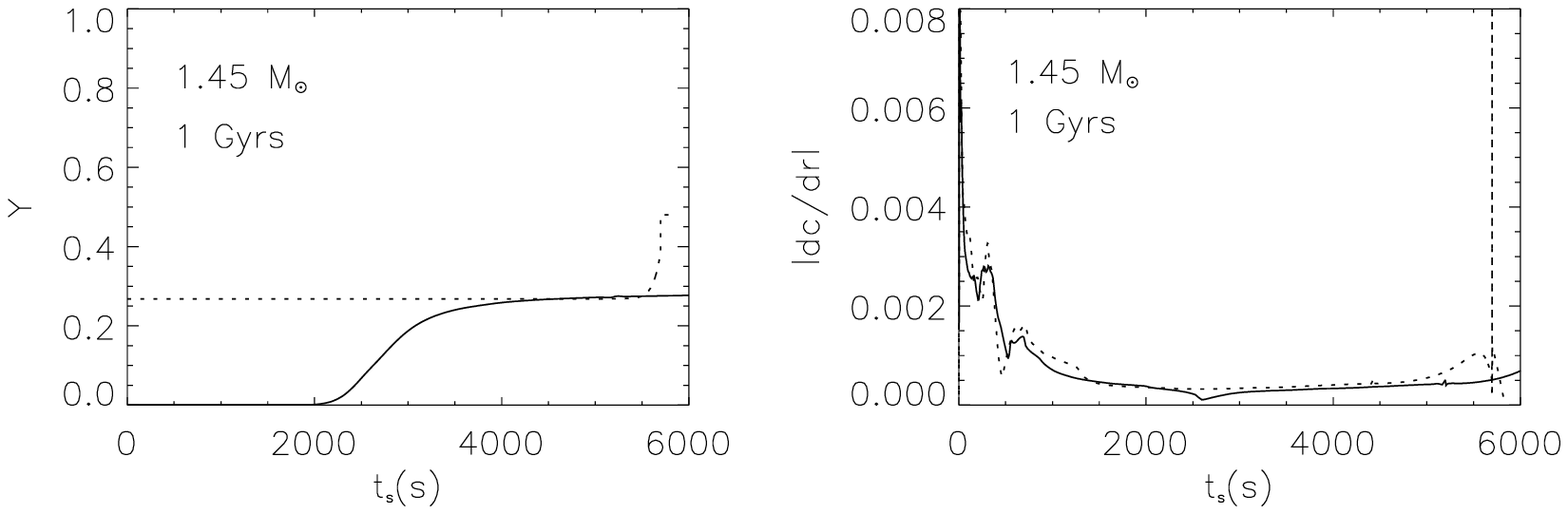}
\includegraphics[angle=0,height=4cm,width=\columnwidth]{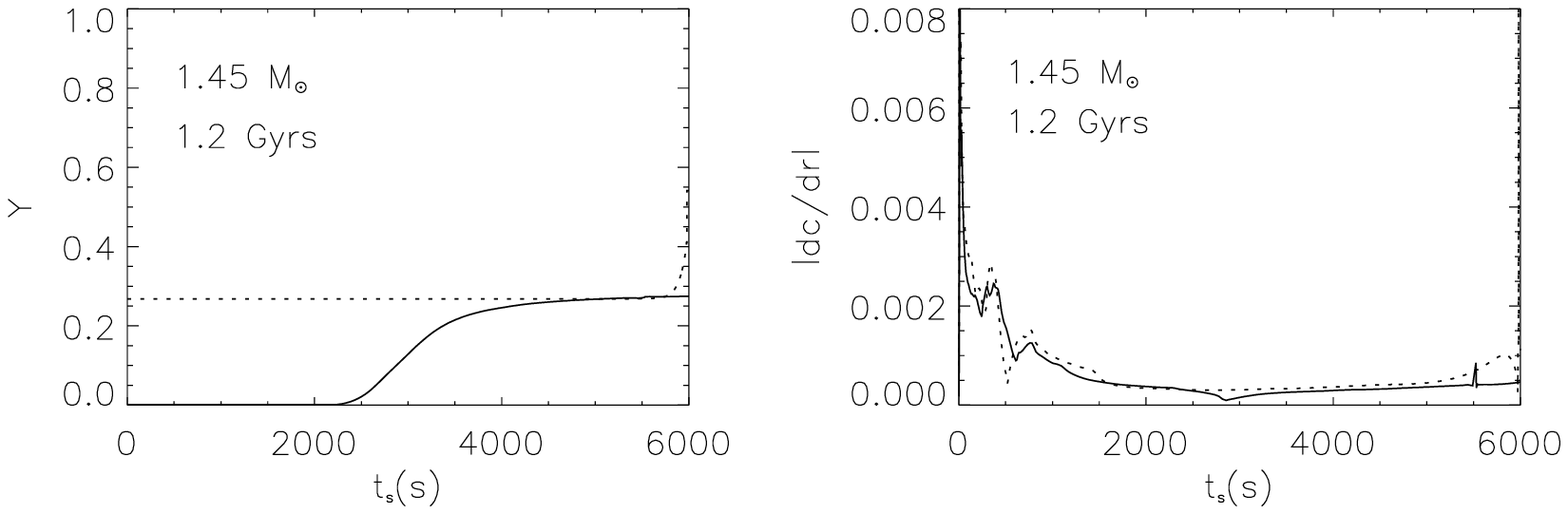}
\includegraphics[angle=0,height=4cm,width=\columnwidth]{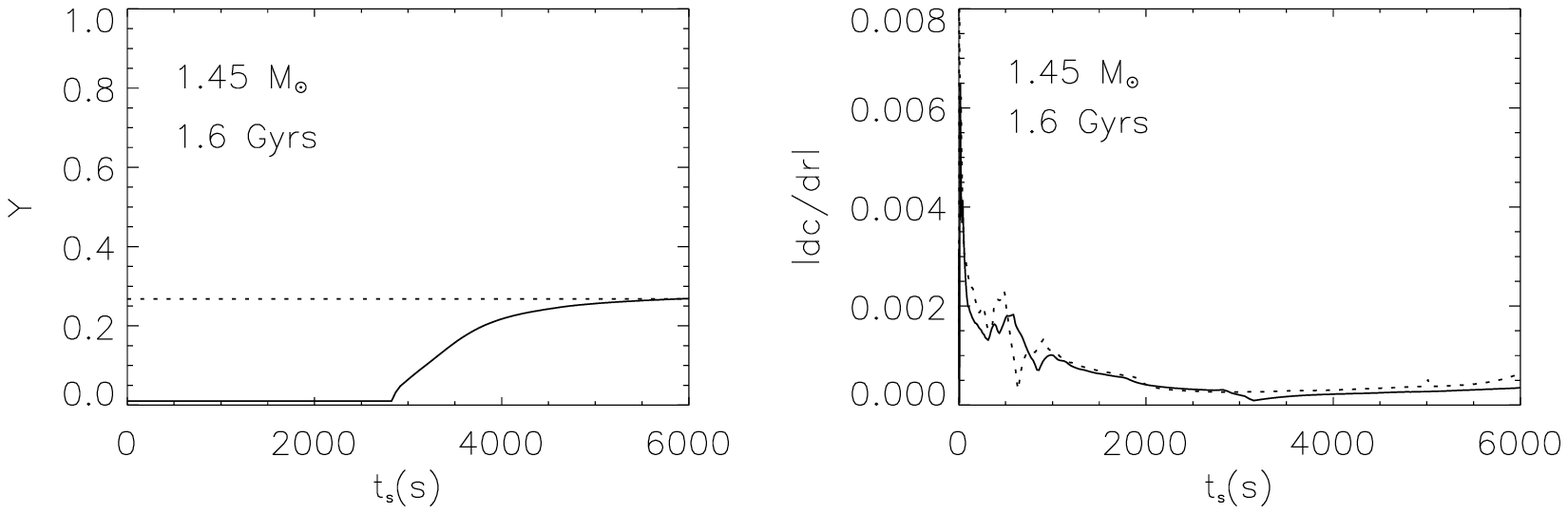}
\includegraphics[angle=0,height=4cm,width=\columnwidth]{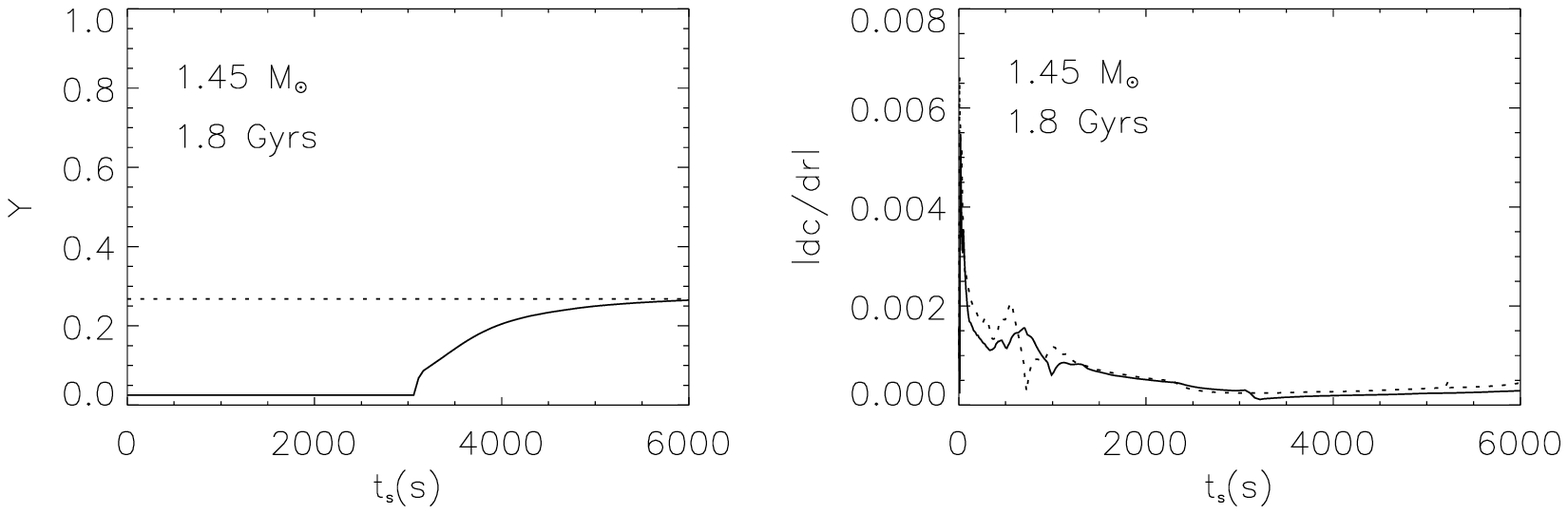}
\end{center}
\caption{Same as Fig. \ref{He1.2} for 1.45 \msol \ at 1, 1.2, 1.6 and 1.8 Gyrs}
\label{He1.45}
\end{figure}

\indent The largest differences between the models with and without diffusion lie just below the convective zone where helium drifts inward due to the effect of gravitation. Figures \ref{He1.2} to \ref{He1.45} display the helium profiles and the gradients of the sound velocity in models with (solid lines) and without (dashed lines) diffusion for M1.2, M1.3 and M1.45, as a function of the acoustic depth (time needed for the waves to travel from the surface down to the considered layer), at different ages. Models M1.1 behave in a way similar to models M1.2 and models M1.35 are similar to M1.3 : they are available in the electronic version only. In the helium profiles, helium gradients due to diffusion are clearly seen. In models M1.1 and M1.2, this helium gradient remains smooth throughout the evolution in the main sequence, while in the case of models M1.3 to M1.45, it becomes sharper with time. For example, in models M1.3, the bottom of the convective zone deepens in the stellar interior around 2.5 Gyrs, which explains the discontinuity on the helium gradient and the increase of the helium mass fraction in Table \ref{tab:models} for the corresponding model, while in models M1.2, the convective zone sinks later and slower. Actually, the helium gradients in models M1.1 and M1.2 become steeper after the turnoff. In the later (Figure \ref{He1.2}), the features around 700 s in the sound speed profile are due to the helium ionisation zones while the bottom of the convective zone leads to characteristic features at t=1900 s (1 Gyrs), t=2000 s (1.5 Gyrs), t=2200s (2.6 Gyrs) and t=2500 s (3.3 Gyrs). We can see in this example that the differences between the models with and without diffusion increase with time, and that this behaviour increases with increasing stellar mass.

\subsection{The second differences}

\begin{figure}[h!]
\begin{center}
\includegraphics[angle=0,height=4cm,width=\columnwidth]{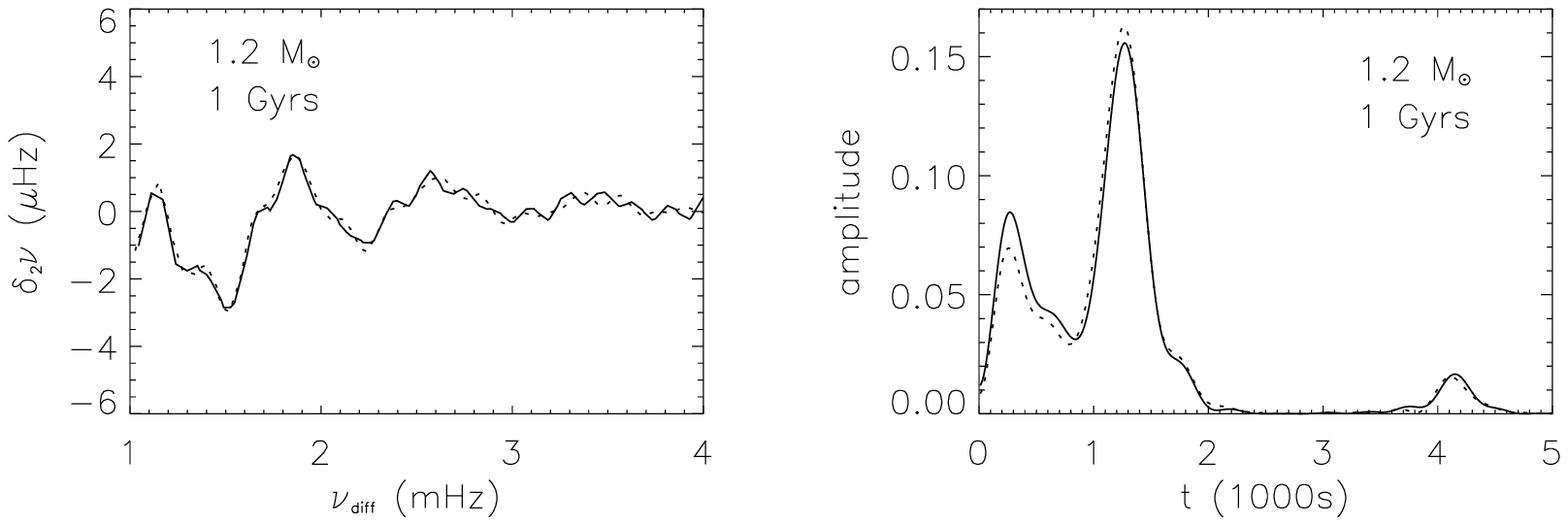} \\
\includegraphics[angle=0,height=4cm,width=\columnwidth]{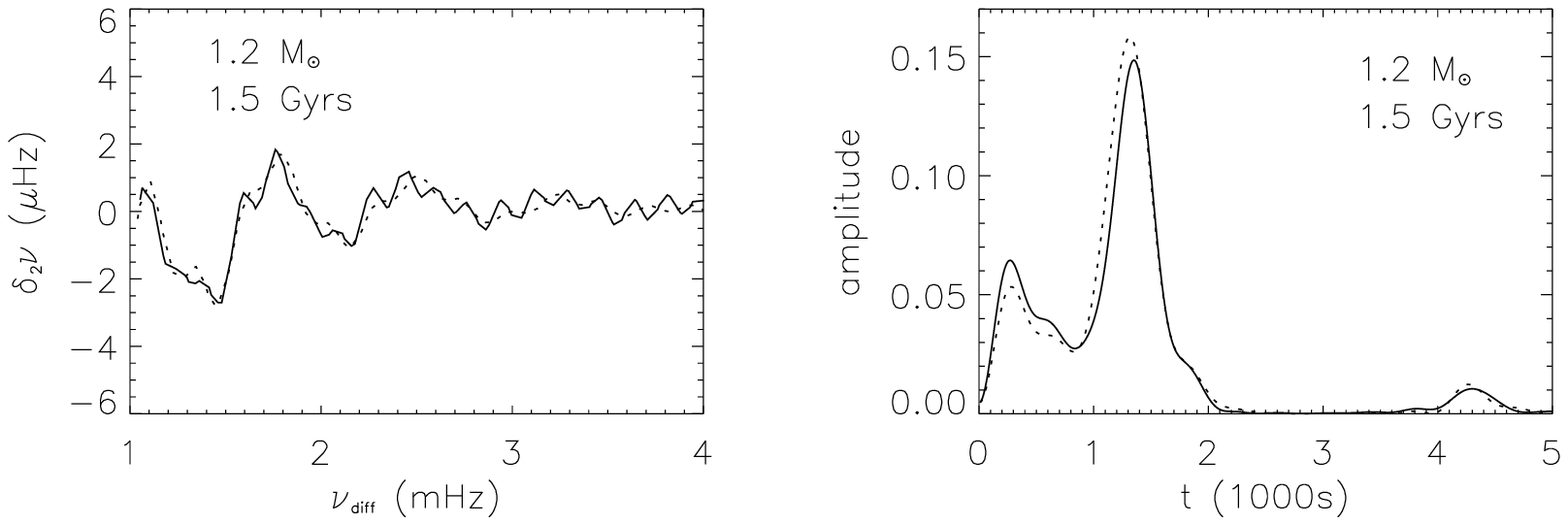} \\
\includegraphics[angle=0,height=4cm,width=\columnwidth]{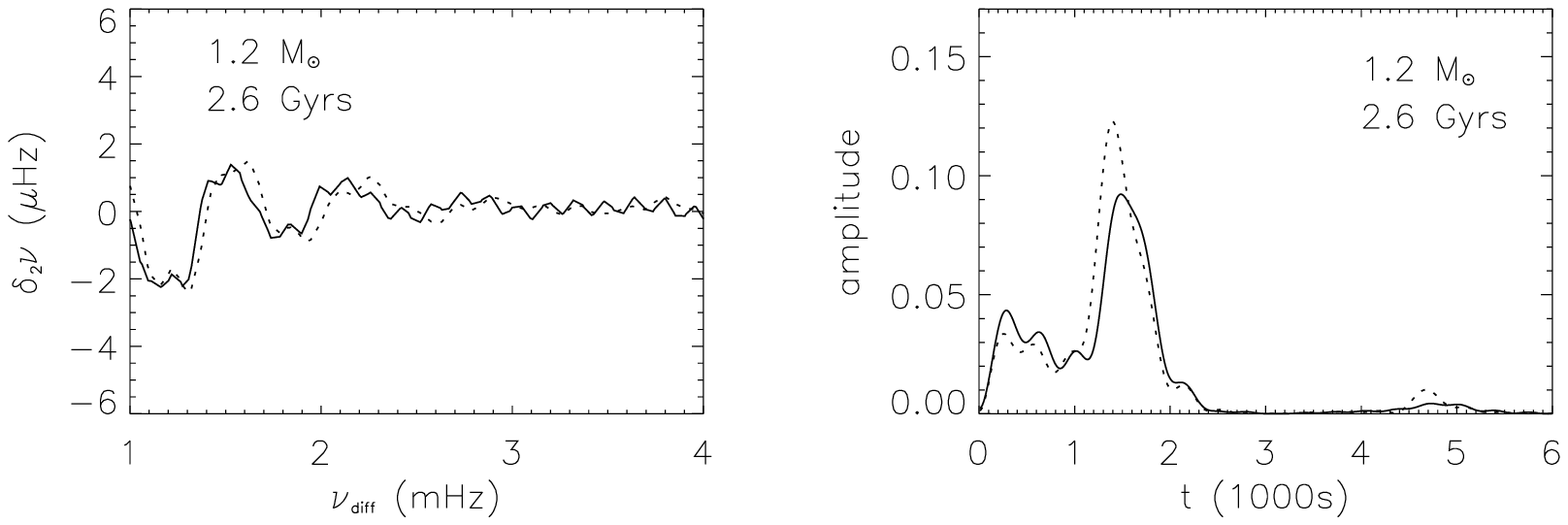} \\
\includegraphics[angle=0,height=4cm,width=\columnwidth]{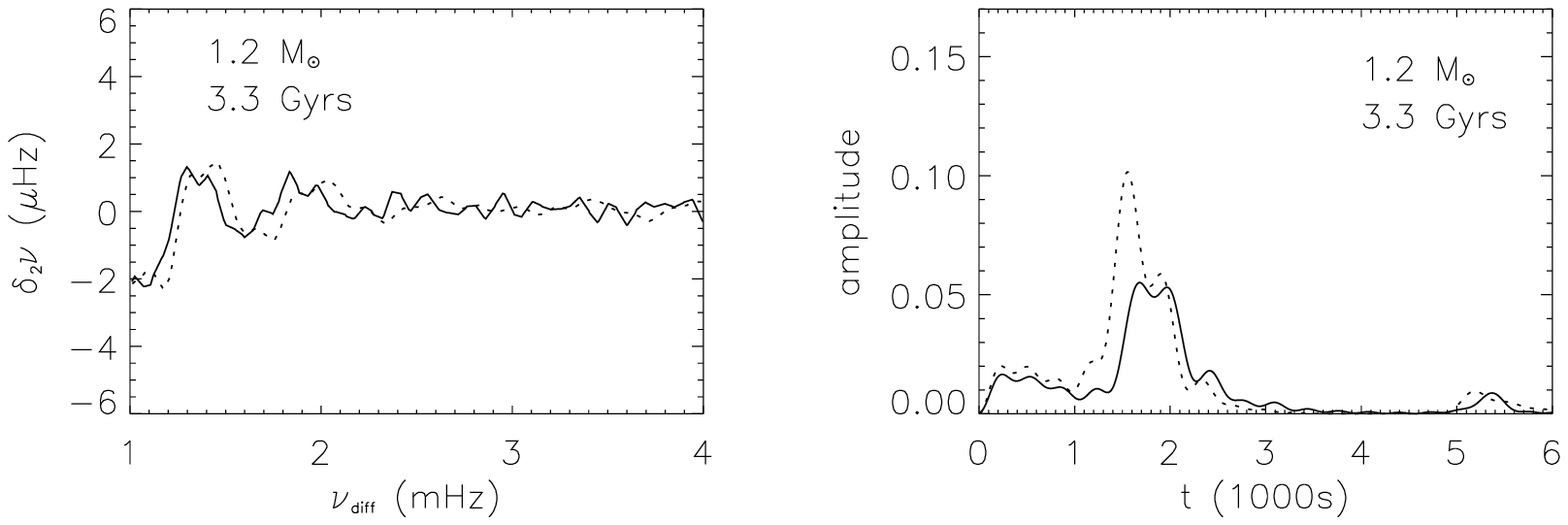}
\end{center}
\caption{Second differences (left column), Fourier transform of the second differences (right column) of models M1.2 of 1.2 \msol \ at 1, 1.5, 2.6 and 3.3 Gyrs, with (solid lines) and without (dashed lines) diffusion.}
\label{astero12}
\end{figure}

\begin{figure}[h!]
\begin{center}
\includegraphics[angle=0,height=4cm,width=\columnwidth]{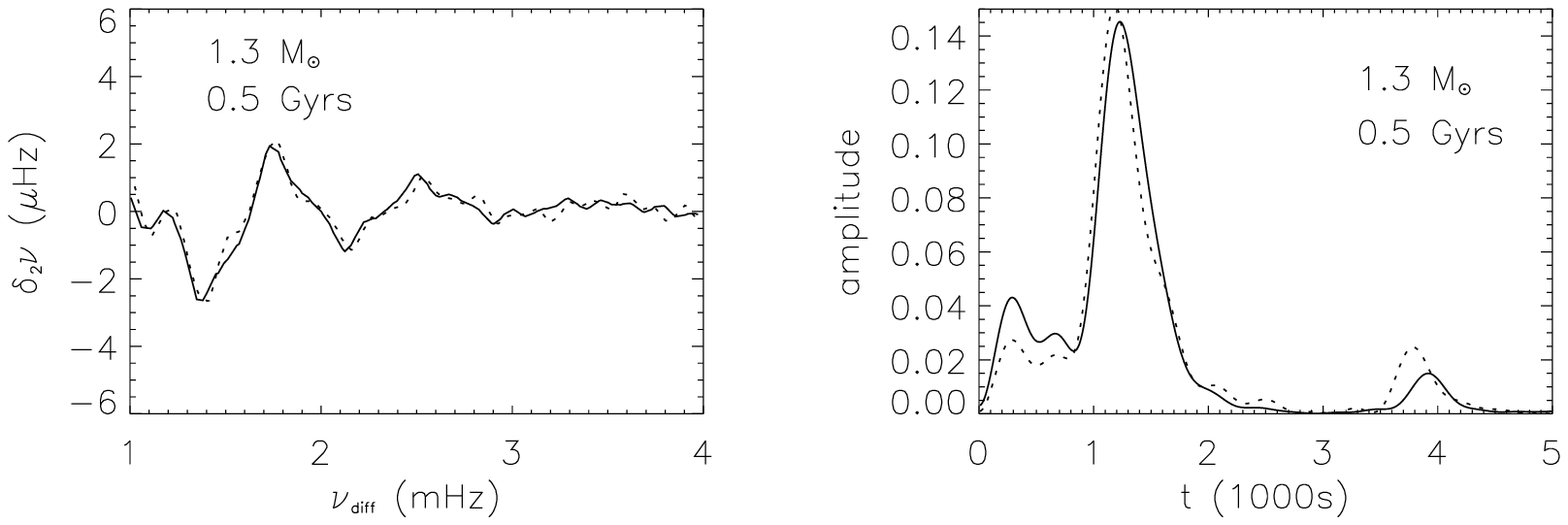} \\
\includegraphics[angle=0,height=4cm,width=\columnwidth]{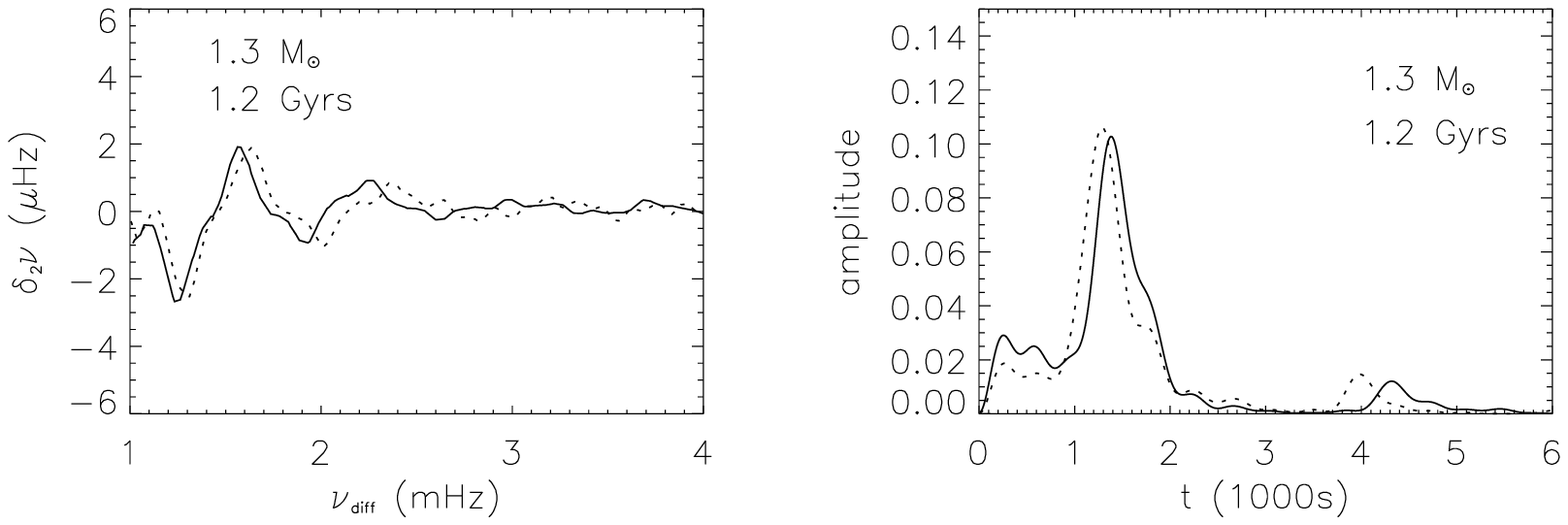} \\
\includegraphics[angle=0,height=4cm,width=\columnwidth]{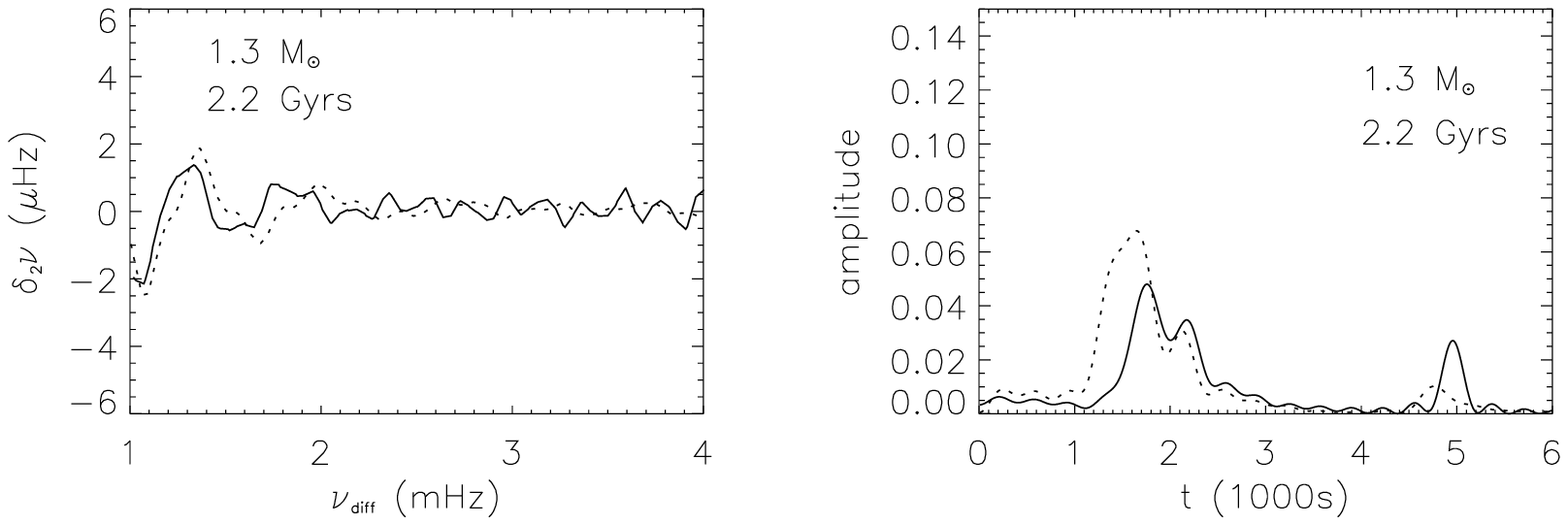} \\
\includegraphics[angle=0,height=4cm,width=\columnwidth]{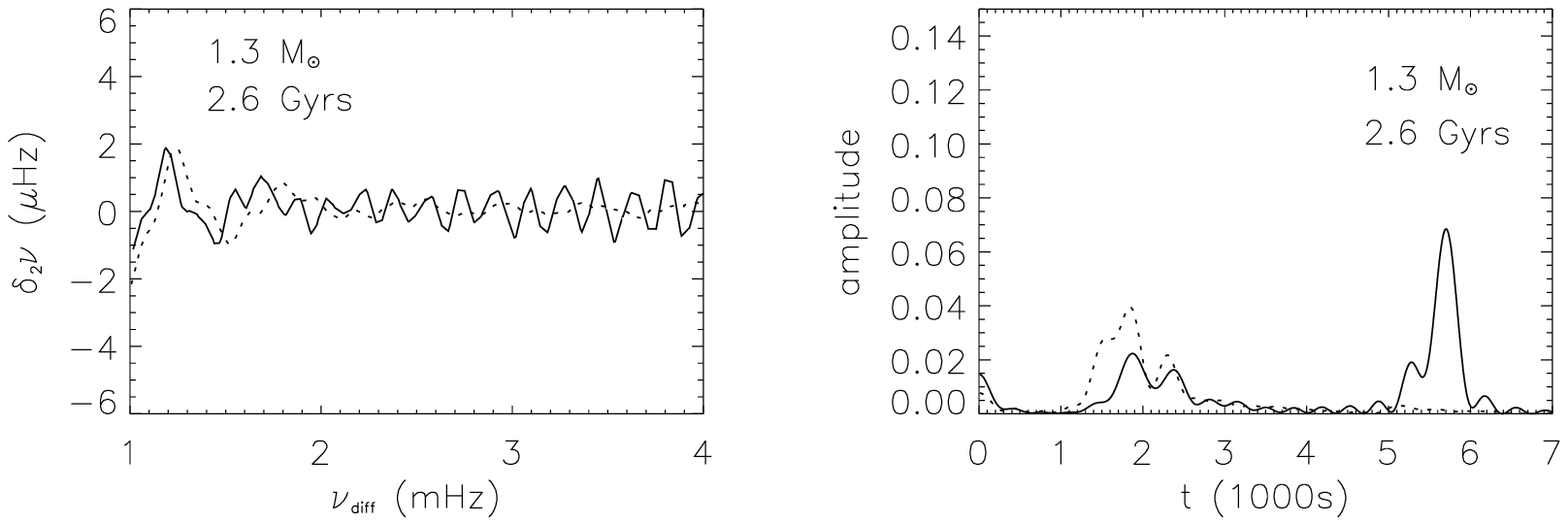}
\end{center}
\caption{Same as Fig. \ref{astero12} for models M1.3 of 1.3 \msol  \ at 0.5, 1.2, 2.2 and 2.6 Gyrs}
\label{astero13}
\end{figure}

\begin{figure}[h!]
\begin{center}
\includegraphics[angle=0,height=4cm,width=\columnwidth]{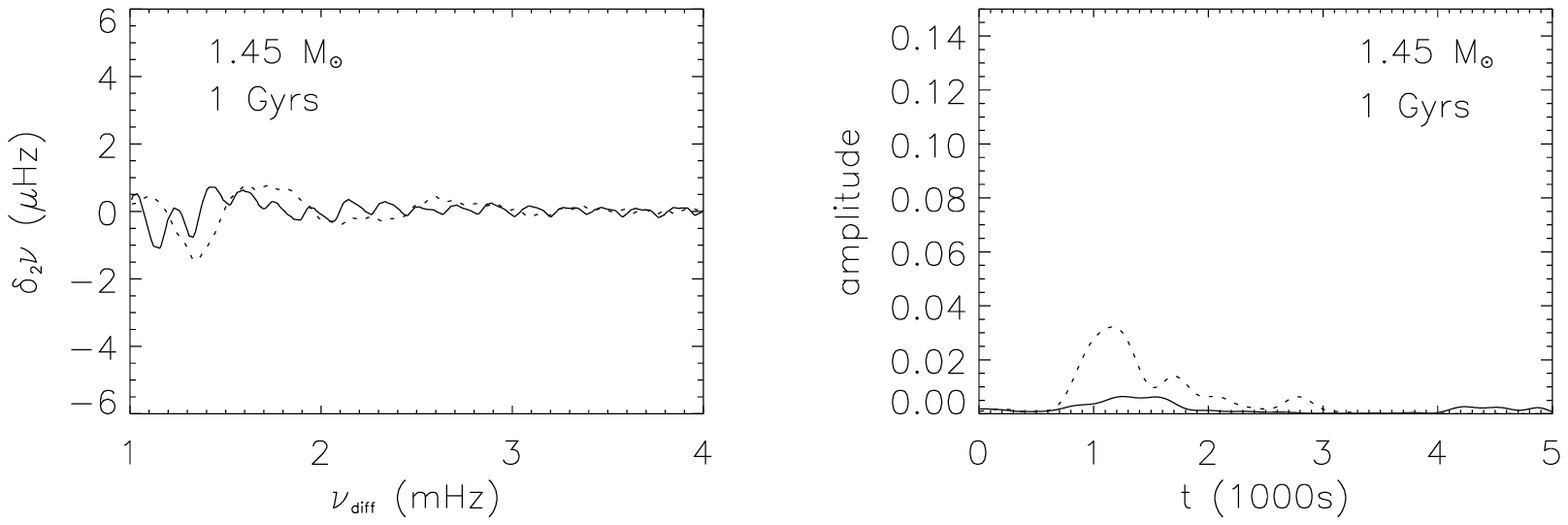} \\
\includegraphics[angle=0,height=4cm,width=\columnwidth]{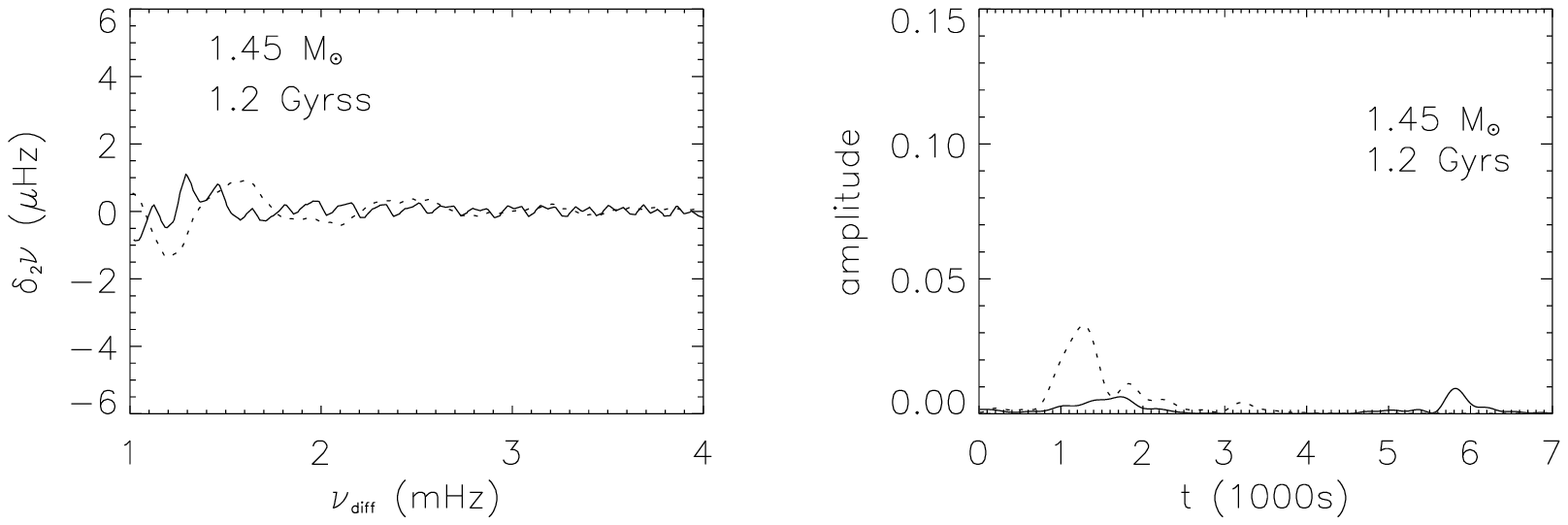} \\
\includegraphics[angle=0,height=4cm,width=\columnwidth]{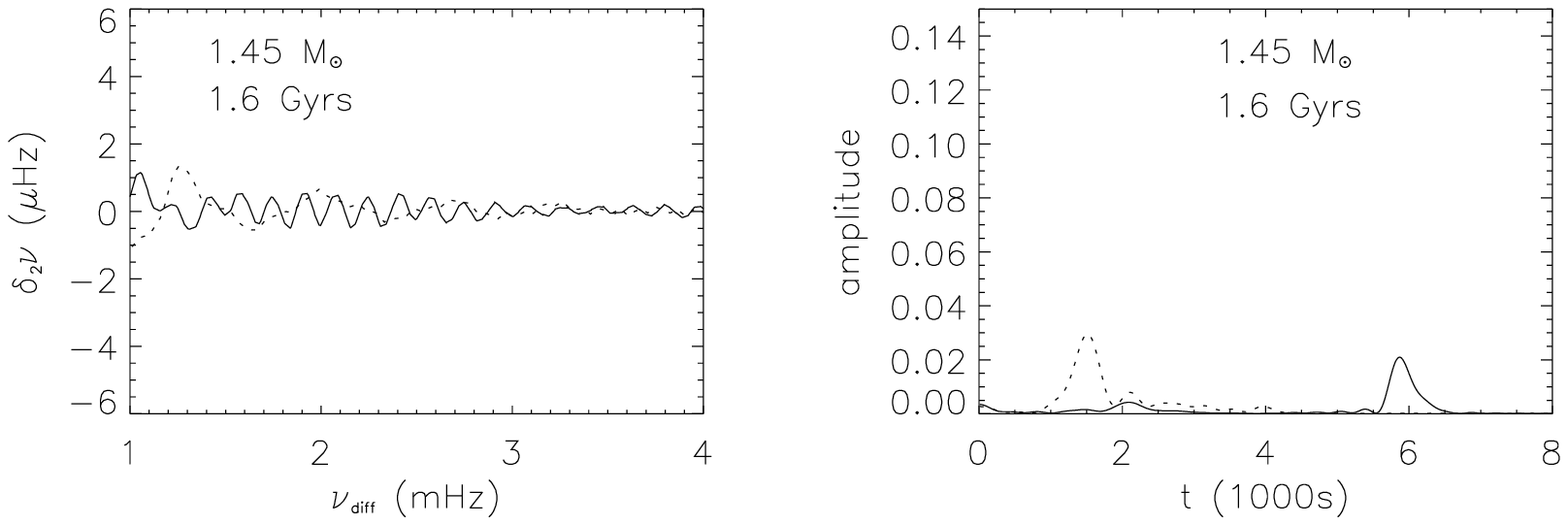} \\
\includegraphics[angle=0,height=4cm,width=\columnwidth]{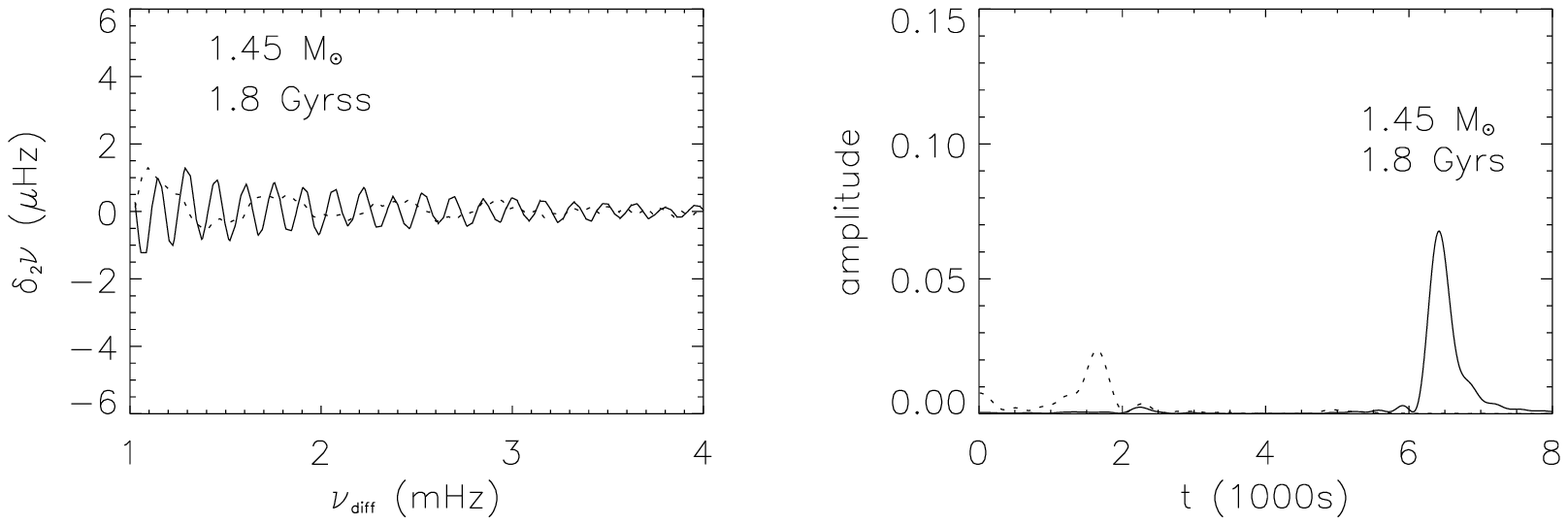}
\end{center}
\caption{Same as Fig. \ref{astero12} for models M1.45 of 1.45 \msol  \ at 1, 1.2, 1.8 and 2.2 Gyrs}
\label{astero145}
\end{figure}

Asteroseismology provides tools to probe the base of the convective zone and the helium gradient. As we have seen, in these locations the sound velocity undergoes rapid variations, which involve partial reflections of the sound waves. These reflections modulate the computed frequencies of the stellar spectrum, which is better seen in the so-called ``second differences'' \citep{gough90,monteiro98,RV01,vauclair04,theado05}. The second differences are defined as follows:
\begin{equation}
\delta_2 \nu = \nu_{n+1} + \nu_{n-1} - 2\nu_n.
\end{equation}
\indent  In order to identify the different components which modulate the oscillations, we have computed the Fourier transform of the second differences. The positions of the peaks correspond to the modulation periods of the second differences due to the partial reflections of the waves, which are equal to twice the acoustic depth of the corresponding features in the sound velocity \citep{vauclair04}. In Figures \ref{astero12} to \ref{astero145} we can then identify for each model the peak due to the helium ionisation zone (the closest to the surface) and those due to the bottom of the convective zone, where the helium gradient lies. The differences between models with and without diffusion are clearly visible.

In all the cases with diffusion (Figures \ref{astero12} to \ref{astero145}), the peaks due to the helium ionisation zones decrease in amplitude during evolution, because of the helium concentration decreases in the convective zone due to diffusion.

In the 1.1 \msol and 1.2 \msol \ stars, the diffusion process works on a time scale too long to be visible in the helium distribution during the main sequence. The amplitude of the peak due to the base of the convective zone slightly decreases because of the sinking of this zone (see Figure \ref{astero12}). For higher mass stars with diffusion (Figures \ref{astero13} and \ref{astero145}), the peak due to the base of the convective zone undergoes a strong increase of amplitude when the convective zone becomes deeper and steepens the helium gradient. In the homogeneous models, there is no increase in the amplitude of this peak. This behaviour is clearly due to the diffusion process.

\section{Discussion}

The large and small separations represent asteroseismic tests of the different internal structure in models with and without diffusion. Our computations show that these differences increase during stellar evolution and that it becomes greater for increasing mass. This behaviour should be detectable for stellar masses larger than 1.2 \msol. 

The behaviour of the second differences of the models are clear : due to the effect of diffusion, helium falls below the outer convective zone and forms a gradient that increases during the evolution. But, depending on the mass of the star, the behaviour of the sound waves is different. For low masses (1.1 and 1.2 \msol), the diffusion process is slow, and the depth of the convective zone increases slowly. Hence, the helium gradient remains smooth during the main sequence and becomes steep only at the turnoff. When the mass of the star is large enough, the convective zone deepens more quickly and the helium gradient steepens more rapidly. There is a strong reflection of the pressure waves in the region of the helium gradient, which explains the large increase in the amplitude of the peak related to the base of the convective zone. Later, helium dilutes inside the convective zone, so that the helium concentration, which was very low after rapid diffusion, increases again (see Table \ref{tab:models}). However, the helium gradient remains steep and the amplitude of the peak remains high (see Figures \ref{astero13} and \ref{astero145}).

The models presented here have been computed with pure gravitational settling. In real stars, the variation in the helium abundance and the steepening of the helium gradient resulting from diffusion will be smaller, due to the mixing processes below the convective zone and possible mass loss. We can however expect that such a signature of helium diffusion may be detected in stars slightly more massive than the Sun with CoRoT. The accuracy of the instrument is expected to reach 0.1 $\mathrm{\mu}$Hz and should allow detection of peaks in the Fourier transforms of the second differences of the oscillation frequencies as a signature of helium diffusion.
  
\bibliographystyle{aa}

\end{document}